\documentclass[reqno, eucal]{amsart}

\usepackage[mathscr]{eucal} %  use \EuScript (\mathcal unchanged)
\usepackage[dvips]{graphicx}
\usepackage[dvips]{color}
\usepackage{amsmath,amsfonts,amssymb,amsthm,amscd}
\usepackage{epic}
\usepackage{eepic}
\usepackage{longtable}
\usepackage{array}
\usepackage{here}

\numberwithin{equation}{section}

\newtheorem{theorem}{Theorem}[section]

\newtheorem{conjecture}[theorem]{Conjecture}
\newtheorem{proposition}[theorem]{Proposition}
\newtheorem{corollary}[theorem]{Corollary}

\theoremstyle{definition}

\newtheorem{example}[theorem]{Example}
\newtheorem{remark}[theorem]{Remark}

\newcommand{\Z}{{\mathbb Z}}

\newcommand{\C}{{\mathbb C}}

\newcommand{\kt}{{\rangle}}

\begin{document}

\title[Multispecies TASEP]{Multispecies TASEP and combinatorial 
${\boldsymbol R}$}

\author{Atsuo Kuniba}
\email{atsuo@gokutan.c.u-tokyo.ac.jp}
\address{Institute of Physics, University of Tokyo, Komaba, Tokyo 153-8902, Japan}

\author{Shouya Maruyama}
\email{maruyama@gokutan.c.u-tokyo.ac.jp}
\address{Institute of Physics, University of Tokyo, Komaba, Tokyo 153-8902, Japan}

\author{Masato Okado}
\email{okado@sci.osaka-cu.ac.jp}
\address{Department of Mathematics, Osaka City University, 
3-3-138, Sugimoto, Sumiyoshi-ku, Osaka, 558-8585, Japan}

\maketitle

\begin{center}
{\it Dedicated to the memory of Professor Ryogo Hirota$\qquad\quad\;$}
\end{center}

\vspace{0.5cm}
\begin{center}{\bf Abstract}
\end{center}

We identify the algorithm for constructing steady states of 
the $n$-species totally asymmetric simple exclusion process
(TASEP) on $L$ site periodic chain by Ferrari and Martin 
with a composition of combinatorial $R$ for the quantum affine algebra  
$U_q(\widehat{sl}_L)$ in crystal base theory.
Based on this connection and the factorized form of the 
$R$ matrix derived recently from the tetrahedron equation,  
we establish a new matrix product formula for the steady state of the TASEP
which is expressed in terms of corner transfer matrices of the 
$q$-oscillator valued five-vertex model at $q=0$.
 
\vspace{0.4cm}

\section{Introduction}\label{sec:intro}

In this paper and the next \cite{KMO} we launch a new approach 
and integrable structure on 
the $(n+1)$-state totally asymmetric simple exclusion process 
on a one-dimensional (1D) periodic chain $\Z_L$ of $L$ sites.
It is called  $n$-species TASEP or $n$-TASEP for short.
A local spin at each site takes values in $\{0,1,\ldots, n\}$
and the neighboring pairs $(\alpha, \beta)$ with 
$\alpha>\beta$ are exchanged to $(\beta, \alpha)$ with a 
uniform probability.
It is a model of non-equilibrium stochastic dynamics 
in physical, biological and many other systems including traffic flow, etc.
See for example \cite{DEHP, BE} and reference therein.

The first basic problem in the $n$-TASEP is to determine its steady state,
which exhibits an intriguing non-uniform measure for $n \ge 2$ 
as in Example \ref{ex:pbar}.
It has been solved in \cite{FM} by introducing a companion system on 
an $n$-tuple of $\{0,1\}$-sequences called {\em multiline process}.
Steady states of the multiline process possess a uniform measure and 
those for the $n$-TASEP are obtained as the image by a certain projection $\pi$ 
from the former to the latter.
The map $\pi$ playing the key role is 
constructed via a combinatorial procedure on the 
$n$-tuple of $\{0,1\}$-sequences which we call the Ferrari-Martin algorithm.

Our first main observation is that the Ferrari-Martin algorithm is most 
naturally formulated in terms of {\em combinatorial $R$} in crystal base theory,
a theory of quantum group at $q=0$ \cite{Ka1}.
More specifically, each $\{0,1\}$-sequence in the multiline process 
is regarded as a crystal $B^l$ (\ref{VBdef}) 
of an antisymmetric tensor representation of 
the Drinfeld-Jimbo quantum affine algebra $U_q(\widehat{sl}_L)$.
The ``2-body operation" of the adjacent $\{0,1\}$-sequences
described as arrival/service/departure \cite{FM} is nothing but the 
combinatorial $R$ acting on the tensor product 
$B^l\otimes B^m$ of crystals \cite{NY}.
It is the bijection that arises from the quantum $R$ matrix
at $q=0$ (crystallization)
which still satisfies the Yang-Baxter equation.
The ``$n$-body operation" in the multiline process is identified, 
by using the Yang-Baxter equation,
with the {\em corner transfer matrix} \cite{Bax}
of the size-$n$ vertex model whose Boltzmann weights are 
crystallized to the combinatorial $R$. 
See (\ref{ctm1}).
We remark that such a size-$n$ system equipped with 
the internal symmetry $U_q(\widehat{sl}_L)$ 
corresponds to the cross-channel of the original $n$-TASEP in the sense that 
the role of the physical space $\Z_L$ 
and the internal space $\{0,\ldots, n\}$ is interchanged.
In particular the cyclic symmetry of the physical space $\Z_L$ in the 
$n$-TASEP has been 
incorporated into the Dynkin diagram of $U_q(\widehat{sl}_L)$.
The integrability in the direct-channel, i.e. 
a relation to a $U_q(\widehat{sl}_{n+1})$ spin chain on $\Z_L$,
has been demonstrated for a more general ASEP in \cite[Sec.5]{AKSS}.

Our second result is the matrix product representation 
(\ref{ptr}) of the steady state probability.
It is expressed in terms of the operator $X_i$ (\ref{Xn})  which can be regarded as
a corner transfer matrix of the 
five-vertex model (\ref{5v}) whose Boltzmann weights 
take values in a $q$-deformed oscillator algebra 
at $q=0$ (\ref{ac02}).
It is obtained by combining the crystal formulation of the Ferrari-Martin algorithm 
with the factorized form of the combinatorial $R$ (\ref{RLpic}).
The latter is the $q=0$ corollary of the factorized form of
the quantum $R$ matrix established recently \cite{KOS}. 
The origin of the factorization itself goes further back to 
a 3D generalization of the Yang-Baxter equation 
known as the {\em tetrahedron equation} \cite{Zam80}.
In fact our operator $X_i$, which is 
different from the earlier one \cite{EFM},
possesses a far-reaching generalization 
in the framework of a 3D lattice model associated with 
the tetrahedron equation \cite{KMO}.

The layout of the paper is as follows.
In Section \ref{sec:uq} we recall the background from 
$U_{q}(\widehat{sl}_L)$ and the combinatorial $R$ 
necessary in this paper.
In Section \ref{sec:tasep}  we treat the $n$-TASEP and 
reformulate the Ferrari-Martin algorithm 
in terms of crystals and combinatorial $R$.
In Section \ref{sec:mpa} a new matrix product formula for the 
steady state probability is derived by synthesizing the contents in 
Section \ref{sec:uq} and Section \ref{sec:tasep}.
Section \ref{sec:dis} contains a remark on a
generalization of stochastic dynamics from the viewpoint of 
crystal base theory and a brief announcement on the so called hat relation.
There are notable 3D structures connected to the tetrahedron equation   
behind several scenes in this paper.
They will be fully demonstrated in our forthcoming paper \cite{KMO}.

\section{Background facts from $U_{q}(\widehat{sl}_L)$}\label{sec:uq}
Let us recapitulate relevant results from the 
representation theory of quantum affine algebras, 
matrix product forms of the quantum $R$ matrix and combinatorial $R$.

\subsection{\mathversion{bold}$U_{q}(\widehat{sl}_L)$ 
and antisymmetric representation}
\label{ss:uq}
The Drinfeld-Jimbo quantum affine algebra 
$U_q=U_{q}(\widehat{sl}_L)$ is a Hopf algebra 
with generators $e_i, f_i, k^{\pm 1}_i\,(i \in \Z_L)$ \cite{D86,J1}
satisfying the Serre relation and 
\begin{align*}
&k_i k^{-1}_i = k^{-1}_ik_i = 1,\; [k_i, k_j]=0,\;
k_i e_j = q^{a_{i,j}}e_j k_i,\; k_i f_j = q^{-a_{i,j}}f_j k_i,\;
[e_i,f_j] = \delta_{i,j}\frac{k_i-k^{-1}_i}{q-q^{-1}},\\
&e^2_ie_j -(q+q^{-1})e_ie_je_i + e_je^2_i = 0,\;
f^2_if_j -(q+q^{-1})f_if_jf_i + f_jf^2_i = 0\;\;(i-j \equiv \pm 1\,\mathrm{mod}\, L).
\end{align*}
Here $a_{i,j} = 2\delta_{i,j} - \delta_{i,j-1}-\delta_{i,j+1}$
with $\delta_{i,j} = 1$ if 
$i\equiv j\; \mathrm{mod}\,L$ and $\delta_{i,j} =0$ otherwise.
All the indices are to be understood in $\Z_L$ likewise.
We adopt the coproduct $\Delta: U_q\rightarrow U_q \otimes U_q$ of the form
\begin{align*}
\Delta e_i = e_i \otimes 1 + k_i \otimes e_i,\quad
\Delta f_i = 1 \otimes f_i + f_i \otimes k^{-1}_i,\quad 
\Delta k^{\pm 1}_i = k^{\pm 1}_i \otimes k^{\pm 1}_i.
\end{align*}
Our $\Delta$ here is opposite of \cite{KOS}.
We assume that $q$ is a generic complex parameter 
in Section \ref{ss:uq}--\ref{ss:mpar}.

We are concerned with the degree $l$ antisymmetric tensor representation 
$\phi_x:  U_{q}(\widehat{sl}_L) \rightarrow \mathrm{End} \,V^l$ with 
$1 \le l < L$. The representation space $V^l$ is given by
\begin{align}\label{VBdef}
V^l = \bigoplus_{{\bf b} \in B^l} \C |{\bf b}\rangle,\qquad
B^l = \{{\bf b}=(b_1,\ldots, b_L)  \in \{0,1\}^L
\mid |{\bf b}| = l\},
\end{align}
where $|{\bf b}| =b_1+ \cdots + b_L$.
Thus $\dim V^l = \binom{L}{l}$.
The labeling set $B^l$ of the basis of $V^l$
is called the {\em crystal}  of $V^l$.
Note that the dependence on $L$ is not indicated in the notation $B^l$.
To save the space, elements of $B^l$ 
will often be written as words on $\{0,1\}$, .e.g.
$(1,0,1,1,0) \in B^3$ is just denoted by $10110$, etc\footnote{A more 
standard notation is a one-column
semi-standard tableau, the one filled with $1,3,4$ for this example.}.

Generators act on $V^l$ as 
($\phi_x(g)$ for a generator $g$ is denoted by $g$ for simplicity)
\begin{align*}
e_i|{\bf b}\rangle = 
x^{\delta_{i,0}}|{\bf b}+{\bf e}_i-{\bf e}_{i+1}\rangle,\;
f_i|{\bf b}\rangle = 
x^{-\delta_{i,0}}|{\bf b}-{\bf e}_i+{\bf e}_{i+1}\rangle,\;
k^{\pm 1}_i|{\bf b}\rangle = 
q^{\pm(b_i-b_{i+1})}|{\bf b}\rangle,
\end{align*}
where ${\bf e}_i = (0,\ldots, 0,\overset{i}{1},0,\ldots, 0)
\in \Z^L$ and the right hand sides are to be understood as $0$ unless 
the arrays in the ket stay within $\{0,1\}^L$.
The representation $\phi_x$ is irreducible and $x$ is called a spectral parameter.

\subsection{\mathversion{bold}Quantum $R$ matrix}
Let $\Delta_{x,y}= (\phi_x \otimes \phi_y)\Delta$ 
be the tensor product representations on $V^l \otimes V^m$,
where $1 \le l, m <L$ are arbitrary.
It is known that they are irreducible if $z:=x/y$ is generic.
Moreover there is a linear map
${\mathscr R}(z) = {\mathscr R}^{l,m}(z): V^l \otimes V^m \rightarrow V^m \otimes V^l$ 
called a {\em quantum $R$ matrix} which is uniquely characterized by
the intertwining relation with the quantum group 
$\Delta_{x,y}(g){\mathscr R}(z)= {\mathscr R}(z)\Delta_{y,x}(g)$ for 
$g \in U_{q}(\widehat{sl}_L)$
up to an overall scalar.
Let us write its matrix elements on the basis 
$|{\bf i}\rangle \otimes |{\bf j}\rangle \in V^l \otimes V^m$ as 
\begin{align}
{\mathscr R}(z)\bigl(|{\bf i}\rangle \otimes |{\bf j}\rangle\bigr)
= \sum_{{\bf a},{\bf b}}
{\mathscr R}(z)^{{\bf a},{\bf b}}_{{\bf i},{\bf j}}
|{\bf b}\rangle \otimes |{\bf a}\rangle,
\label{sact}
\end{align}
where $(|{\bf i}|,|{\bf j}|) = (l,m)$ and 
the sum extends over ${\bf a, \bf b}$ 
such that $(|{\bf a}|,|{\bf b}|) = (l,m)$.
The matrix elements have the support property reflecting the weight conservation:
\begin{align}
{\mathscr R}(z)^{{\bf a},{\bf b}}_{{\bf i},{\bf j}}= 0 \;\;
\text{unless}\;\; \;{\bf a} + {\bf b} = {\bf i} + {\bf j}\;\; \;( \in \{0,1,2\}^L).
\end{align}
The $R$ matrix  satisfies the Yang-Baxter equation \cite{Bax}
\begin{equation}\label{ybe}
({\mathscr R}^{l,m}(z)\otimes 1)(1\otimes {\mathscr R}^{k,m}(zz'))({\mathscr R}^{k,l}(z')\otimes 1)
= (1\otimes {\mathscr R}^{k,l}(z'))
({\mathscr R}^{k,m}(zz')\otimes 1)(1\otimes {\mathscr R}^{l,m}(z))
\end{equation}
for any $1 \le k,l,m <L$.
We normalize the ${\mathscr R}^{l,m}(z)$ so that 
\begin{align}\label{vro}
{\mathscr R}(z) \bigl(|{\bf e}_{\le l}\rangle \otimes |{\bf e}_{\le m}\rangle\bigr)
= {\bar \varrho}(z)|{\bf e}_{\le m}\rangle \otimes |{\bf e}_{\le l}\rangle,\quad
{\bar \varrho}(z) 
= \!\!\!\!\!\!
\prod_{i=(l+m-L)_+}^{\min(l,m)-1}
\!\!\!\!\!\!\!(1-(-1)^{l+m}q^{l+m-2i}z),
\end{align}
where $(m)_+ = \max(m,0)$ and 
$|{\bf e}_{\le l}\rangle = |{\bf e}_1 + \cdots + {\bf e}_l\rangle \in V^l$.
Then all the elements ${\mathscr R}(z)^{{\bf a},{\bf b}}_{{\bf i},{\bf j}}$ 
are {\em polynomials} in $q$ and $z$.

\begin{example}\label{ex:R11}
The ${\mathscr R}^{1,1}(z)$ is the well known 
$R$ matrix for the vector representation of $U_q(\widehat{sl}_L)$.
The $L=2$ case 
corresponds to the six-vertex model \cite{Bax}.
The nonzero elements read
\begin{align*}
{\mathscr R}(z)^{{\bf e}_i, {\bf e}_i}_{{\bf e}_i, {\bf e}_i}= 1-q^2z,\quad
{\mathscr R}(z)^{{\bf e}_i, {\bf e}_j}_{{\bf e}_i, {\bf e}_j} = q(1-z),
\quad
{\mathscr R}(z)^{{\bf e}_j, {\bf e}_i}_{{\bf e}_i, {\bf e}_j}
= (1-q^2)z^{\theta(i<j)},
\end{align*}
where $1 \le i\neq j \le L$ and $\theta(\text{true}) = 1$, 
$\theta(\text{false}) = 0$.
\end{example}

\begin{example}\label{ex:R12}
Nonzero elements of ${\mathscr R}^{1,2}(z)$ for 
$U_q(\widehat{sl}_3)$ are as follows.
\begin{align*}
&{\mathscr R}^{100,110}_{100,110}=
{\mathscr R}^{010,110}_{010,110}=
{\mathscr R}^{100,101}_{100,101}=
{\mathscr R}^{001,101}_{001,101}=
{\mathscr R}^{010,011}_{010,011}=
{\mathscr R}^{001,011}_{001,011}=1+q^3z,\\
&{\mathscr R}^{001, 110}_{001, 110}
= {\mathscr R}^{010, 101}_{010, 101}
={\mathscr R}^{100, 011}_{100, 011}=q(1+qz),\;\;
{\mathscr R}^{001, 110}_{010, 101}= 
{\mathscr R}^{010, 101}_{100, 011}
=z{\mathscr R}^{100, 011}_{001, 110}=-q(1-q^2)z,\\
&{\mathscr R}^{001, 110}_{100, 011}
= z{\mathscr R}^{010, 101}_{001, 110}
= z{\mathscr R}^{100, 011}_{010, 101}=(1-q^2)z.
\end{align*}
\end{example}

\begin{example}\label{ex:R21}
Nonzero elements of ${\mathscr R}^{2,1}(z)$ 
for $U_q(\widehat{sl}_3)$ are as follows.  
\begin{align*}
&{\mathscr R}^{110,100}_{110,100}=
{\mathscr R}^{110,010}_{110,010}=
{\mathscr R}^{101,100}_{101,100}=
{\mathscr R}^{101,001}_{101,001}=
{\mathscr R}^{011,010}_{011,010}=
{\mathscr R}^{011,001}_{011,001}=1+q^3z,\\
&{\mathscr R}^{011,100}_{011,100}={\mathscr R}^{101,010}_{101,010}
={\mathscr R}^{110,001}_{110,001}=q(1+qz),\;\;
{\mathscr R}^{011,100}_{101,010}={\mathscr R}^{101,010}_{011,001}
=z{\mathscr R}^{110,001}_{011,100}=-q(1-q^2)z,\\
&{\mathscr R}^{011,100}_{110,001}=z{\mathscr R}^{101,010}_{011,100}=
z{\mathscr R}^{110,001}_{101,010}=(1-q^2)z.
\end{align*}
\end{example}
See for example \cite{DO} for more information on the quantum $R$ matrix.

\subsection{\mathversion{bold}Matrix product representation of ${\mathscr R}(z)$}
\label{ss:mpar}
In a recent work \cite{KOS} based on the tetrahedron equation, 
a matrix product representation of the quantum 
$R$ matrix ${\mathscr R}^{l,m}(z)$ was constructed.
Let us quote the result in a form adapted to the present convention.

Consider the $q$-oscillator algebra ${\mathscr A}_q$ generated by 
${\bf a}^+, {\bf a}^-, {\bf k}$ satisfying the relations
\begin{align}
{\bf k}\,{\bf a}^{\pm} = -q^{\pm 1}{\bf a}^{\pm}\,{\bf k},\quad
{\bf a}^+{\bf a}^- = 1-{\bf k}^2,\quad
{\bf a}^-{\bf a}^+ = 1-q^2 {\bf k}^2.
\label{ac2}
\end{align}
They may be identified with the 
operators acting on the Fock space 
$F = \bigoplus_{m \ge 0}\C |m\rangle$\footnote{This ket 
$|m \rangle \in F$ with $m \in \Z_{\ge 0}$ 
should not be confused with 
$|{\bf b}\rangle \in V^l$ with ${\bf b} \in B^l$.} as
\begin{align}
{\bf a}^+|m\rangle = |m+1\rangle,\quad
{\bf a}^-|m\rangle = (1-q^{2m})|m-1\rangle,\quad
{\bf k}|m\rangle = (-q)^m|m\rangle.
\label{ac1}
\end{align}
We arrange them in the {\em 3D $L$ operator}  
${\mathscr L} = ({\mathscr L}^{a,b}_{i,j})$ \cite{BS}  where all the indices 
belong to $\{0,1\}$.
They are zero except the following:
\begin{align}\label{lak}
{\mathscr L}_{0, 0}^{0,0}&= {\mathscr L}_{1,1}^{1,1} = 1,\;\;
{\mathscr L}_{1,0}^{0,1} = {\bf a}^+,\;\;
{\mathscr L}^{1,0}_{0,1} = {\bf a}^-,\;\;
{\mathscr L}_{0,1}^{0,1} = {\bf k},\;\;
{\mathscr L}_{1,0}^{1,0} = q{\bf k}.
\end{align}
The ${\mathscr L}$ may be viewed as defining a 
six-vertex model whose Boltzmann weights belong to ${\mathscr A}_q$, i.e.,
$q$-oscillator valued six-vertex model:
\begin{equation}\label{6v}
\begin{picture}(200,65)(-10,-36)

\put(-60,0){
\put(-52,-3){${\mathscr L}^{a,b}_{i,j}=$}
\put(-10,0){\vector(1,0){20}}
\put(0,-10){\vector(0,1){20}}
\put(-17,-3.5){$i$}\put(12.5,-3){$a$}\put(-2.4,13){$b$}\put(-2.3,-19){$j$}
}

\put(-10,0){\vector(1,0){20}}
\put(0,-10){\vector(0,1){20}}
\put(-17,-3.5){0}\put(12.5,-3.5){0}\put(-2.4,13){0}\put(-2.3,-19){0}
\put(-2,-34){1}

\put(50,0){
\put(-10,0){\vector(1,0){20}}
\put(0,-10){\vector(0,1){20}}
\put(-17,-3.5){1}\put(12.5,-3.5){1}\put(-2.4,13){1}\put(-2.3,-19){1}
\put(-2,-34){1}}

\put(100,0){
\put(-10,0){\vector(1,0){20}}
\put(0,-10){\vector(0,1){20}}
\put(-17,-3.5){1}\put(12.5,-3.5){0}\put(-2.4,13){1}\put(-2.3,-19){0}
\put(-2.5,-34){${\bf a}^+$}}

\put(150,0){
\put(-10,0){\vector(1,0){20}}
\put(0,-10){\vector(0,1){20}}
\put(-17,-3.5){0}\put(12.5,-3.5){1}\put(-2.4,13){0}\put(-2.3,-19){1}
\put(-2.5,-34){${\bf a}^-$}}

\put(200,0){
\put(-10,0){\vector(1,0){20}}
\put(0,-10){\vector(0,1){20}}
\put(-17,-3.5){0}\put(12.5,-3.5){0}\put(-2.4,13){1}\put(-2.3,-19){1}
\put(-2.5,-34){${\bf k}$}}

\put(250,0){
\put(-10,0){\vector(1,0){20}}
\put(0,-10){\vector(0,1){20}}
\put(-17,-3.5){1}\put(12.5,-3.5){1}\put(-2.4,13){0}\put(-2.3,-19){0}
\put(-3.5,-34){$q{\bf k}$}}

\end{picture}
\end{equation}
The relations in (\ref{ac2}) involving 
${\bf a}^\pm{\bf a}^\mp$ are quantization of the 
so called free fermion condition \cite[eq.(10.16.4)]{Bax}$|_{\omega_7=\omega_8=0}$.
One may introduce the third arrow perpendicular to the two
arrows for each vertex encoding the action of the 
corresponding $q$-oscillator generators.
It amounts to regarding $\mathscr{L}$ as a vertex of the 
3D square lattice, hence the name 3D $L$ operator.
The resulting 3D classical spin model 
satisfies the tetrahedron equation \cite[eq.(2.13)]{KOS} involving the 
3D $R$ operator ${\mathscr R}$ obtained by replacing $q$ by $-q$ in 
\cite[eq.(2.2)]{KOS}.  See also \cite{BS}.
Such a 3D structure is a source of the factorization of the 
$R$ matrix in Theorem \ref{th:kos} and ultimately  
the matrix product form of the steady state probability 
that we will derive in Section \ref{sec:mpa}.

Let ${\bf h}$ be the operator on $F$ defined by 
${\bf h}|m \rangle = m|m \rangle$.
Denote by $F^\ast = \bigoplus_{m\ge 0}\C\langle m|$ the 
dual of $F$ with the pairing 
specified by $\langle m|m'\rangle = (q^2)_m\delta_{m,m'}$ with
$(q^2)_m = \prod_{j=1}^m(1-q^{2j})$.
Accordingly we set 
$\mathrm{Tr}(z^{\bf h}X) 
= \sum_{m\ge 0}\frac{z^m\langle m|X|m\rangle}{(q^2)_m}$
for $X \in {\mathscr A}_q$.
Set $\varrho(z) = q^{-(l-m)_+}(1-(-1)^{l+m}q^{|l-m|}z)
{\bar \varrho}(z)$ where ${\bar \varrho}(z)$ is defined in 
(\ref{vro}). 
The following result is a special case $\epsilon_1=\cdots = \epsilon_L=1$ 
of \cite[Th.4.1]{KOS}.

\begin{theorem}\label{th:kos}
Elements of the $R$ matrix (\ref{sact}) 
with the normalization (\ref{vro}) are expressed in the matrix product form
\begin{align}\label{RL}
{\mathscr R}(z)^{{\bf a},{\bf b}}_{{\bf i},{\bf j}}
= \varrho(z)\mathrm{Tr}\bigl(
z^{\bf h}{\mathscr L}^{a_1,b_1}_{i_1, j_1}
\cdots {\mathscr L}^{a_L,b_L}_{i_L, j_L} \bigr),
\end{align}
where ${\bf a}=(a_1,\ldots, a_L)$ and similarly for 
${\bf b}, {\bf i}, {\bf j}$.
\end{theorem}

For instance we have
\begin{align*}
{\mathscr R}^{011,100}_{011,100} = 
\varrho(z)\mathrm{Tr}
\bigl(z^{\bf h}{\mathscr L}^{0,1}_{0,1}{\mathscr L}^{1,0}_{1,0}{\mathscr L}^{1,0}_{1,0}\bigr)
= q^2\varrho(z)\mathrm{Tr}(z^{\bf h}{\bf k}^3) = \frac{q^2\varrho(z)}{1+q^3z}
\end{align*}
in agreement with Example \ref{ex:R21} 
due to $(l,m)=(2,1)$ and $\varrho(z) = q^{-1}(1+qz)(1+q^3z)$.

We remark that 
the operators 
$\epsilon, \delta, A$ used for the $n$-ASEP \cite{PEM} can be realized  
in ${\mathscr A}_{q^{1/2}}$ (\ref{ac2}) via 
$\epsilon \mapsto {\bf a}^+,
\delta \mapsto {\bf a}^-, A \mapsto {\bf k}^2$.

\subsection{\mathversion{bold}Combinatorial $R$}\label{ss:cr}

The product set of crystals $B^l$ and $B^m$ is again called a crystal 
and denoted by $B^l \otimes B^m$.
As mentioned before, all the elements of the quantum $R$ matrix 
${\mathscr R}={\mathscr R}^{l,m}: V^l \otimes V^m \rightarrow 
V^m \otimes V^l$ are polynomials in $q$.
Therefore it makes sense to set $q=0$. 
We thus define 
$R= R^{l,m}={\mathscr R}^{l,m}(z=1)|_{q=0}$\footnote{Although 
$z$ will not be concerned in this paper it is an important ingredient of the  
combinatorial $R$ in the literature.}. 
It turns out that $R$ becomes a 
beautiful bijection among the crystals 
$R^{l,m}: B^l \otimes B^m \rightarrow B^m \otimes B^l$, 
which is called a {\em combinatorial $R$}.

\begin{example}\label{cR12}
Example \ref{ex:R12}  with (\ref{sact}) yields the combinatorial 
$R^{1,2}: B^1 \otimes B^2 \rightarrow B^2 \otimes B^1$ as
\begin{align*}
&100\otimes 110 \mapsto 110 \otimes 100,\quad
100\otimes 101 \mapsto 101 \otimes 100,\quad
100\otimes 011 \mapsto 110 \otimes 001,\\
&010\otimes 110 \mapsto 110 \otimes 010,\quad
010\otimes 101 \mapsto 011 \otimes 100,\quad
010\otimes 011 \mapsto 011 \otimes 010,\\
&001 \otimes 110 \mapsto 101 \otimes 010,\quad
001 \otimes 101 \mapsto 101 \otimes 001,\quad
001 \otimes 011 \mapsto 011 \otimes 001.
\end{align*}
\end{example}

\begin{example}\label{cR21}
Example \ref{ex:R21}  with (\ref{sact}) yields the combinatorial 
$R^{2,1}: B^2 \otimes B^1 \rightarrow B^1 \otimes B^2$ as
\begin{align*}
&110\otimes 100 \mapsto 100 \otimes 110,\quad
110\otimes 010 \mapsto 010 \otimes 110,\quad
110\otimes 001 \mapsto 100 \otimes 011,\\
&101\otimes 100 \mapsto 100 \otimes 101,\quad
101\otimes 010 \mapsto 001 \otimes 110,\quad
101\otimes 001 \mapsto 001 \otimes 101,\\
&011 \otimes 100 \mapsto 010 \otimes 101,\quad
011 \otimes 010 \mapsto 010 \otimes 011,\quad
011 \otimes 001 \mapsto 001 \otimes 011.
\end{align*}
\end{example}

The maps $R^{2,1}$ and $R^{1,2}$ are both bijections inverse to each other.
They are nontrivial in the sense that 
$R ({\bf i}\otimes {\bf j}) \neq {\bf j}\otimes {\bf i}$ in  general.

Back to the general case, 
let $R = R^{l,m}: B^l \otimes B^m \rightarrow B^m \otimes B^l$
be the combinatorial $R$.
Write its matrix element like (\ref{sact}):
\begin{align}
R\bigl({\bf i}\otimes{\bf j}\bigr)
= \sum_{{\bf a},{\bf b}}
R^{{\bf a},{\bf b}}_{{\bf i},{\bf j}}\;
{\bf b}\otimes {\bf a}.\label{sact0}
\end{align}
Then for any ${\bf i}\otimes{\bf j}$, the matrix elements  
$R^{{\bf a},{\bf b}}_{{\bf i},{\bf j}}$ are all $0$ except exactly one 
pair ${\bf b}\otimes {\bf a} \in B^m \otimes B^l$ 
determined from ${\bf i}$ and ${\bf j}$.
An algorithm for finding it, namely the image  
${\bf b} \otimes {\bf a} = R({\bf i}\otimes {\bf j})$, was 
obtained in \cite[Rule 3.10]{NY}\footnote{
Actually we explain the algorithm for its inverse 
since the assumptions $l\le m$ here and $k\ge l$ in \cite{NY} corresponds to 
the opposite situation.}.
We call it NY(Nakayashiki-Yamada)-rule.

\vspace{0.2cm}\noindent
{\em NY-rule for $l\le m$}:

\vspace{0.1cm}\noindent
(i) Given ${\bf i} \otimes {\bf j} \in B^l \otimes B^m$,
draw a tableau for  
${\bf i}$ below and the one for ${\bf j}$ above.
For example,  the tableaux for 
${\bf i} \otimes {\bf j} =1100100100 \otimes 0010111110
\in B^4 \otimes B^6$ look as (i) below
($1$ and $0$ are denoted by $\bullet$ and empty box): 
\[
\begin{picture}(400,49)(-22,-3)

\put(45,38){(i)} \put(174,38){(ii)}\put(303,38){(iii)}

\put(-20,20){${\bf j}=$}
\put(-20,3){${\bf i}=$}

\multiput(0,0)(130,0){3}{
\multiput(0,0)(0,18){2}{
\put(0,10){\line(1,0){100}}
\multiput(0,0)(10,0){11}{\line(0,1){10}}
\put(0,0){\line(1,0){100}}
}}

\multiput(0,-0.1)(130,0){2}{
\put(23,20){$\bullet$}\put(43,20){$\bullet$}\put(53,20){$\bullet$}
\put(63,20){$\bullet$}\put(73,20){$\bullet$}\put(83,20){$\bullet$}

\put(3,3){$\bullet$}\put(13,3){$\bullet$}
\put(43,3){$\bullet$}\put(73,3){$\bullet$}
}

\put(260,-0.1){
\put(43,20){$\bullet$}
\put(63,20){$\bullet$}\put(73,20){$\bullet$}\put(83,20){$\bullet$}

\put(3,3){$\bullet$}\put(13,3){$\bullet$}\put(23,3){$\bullet$}
\put(43,3){$\bullet$}\put(53,3){$\bullet$}\put(73,3){$\bullet$}
}

\put(130,-0.1){

\put(6,5){\line(0,1){7.9}}\put(6.1,12.7){\line(-1,0){6}}
\multiput(0,0)(-4,0){3}{\put(-1.2,12.7){\line(-1,0){1.8}}}

\put(0,0.2){
\put(15.8,5){\line(0,1){10}}\put(15.9,14.8){\line(-1,0){15.8}}
\multiput(0,2.1)(-4,0){3}{\put(-1.2,12.7){\line(-1,0){1.8}}}
}

\put(45.5,6){\line(0,1){17}}

\put(-0.2,-0.6){
\put(75.5,6.5){\line(0,1){6.5}}\put(75.5,13){\line(-1,0){10}}
\put(65.5,13){\line(0,1){8}}
}

\put(0.2,0){
\put(75.5,23){\line(0,-1){9.4}}\put(75.5,13.6){\line(1,0){25}}
\multiput(100.5,1)(4,0){3}{\put(1.2,12.6){\line(1,0){1.8}}}
}

\put(85.3,23){\line(0,-1){7.4}}\put(85.3,15.6){\line(1,0){15}}
\multiput(100.5,2.9)(4,0){3}{\put(1.2,12.7){\line(1,0){1.8}}}

}

\put(365,20){$={\bf a}$}
\put(365,3){$={\bf b}$}

\end{picture}
\]

\noindent
(ii-1) Pick any dot (call it $d$) in ${\bf i}$  
and connect it to a dot (call it $d'$) in ${\bf j}$.
The partner $d'$ must be the rightmost one among the dots located  
weakly 
left\footnote{``Weakly left" means exactly above or strictly left. 
``Weakly right" means similar.} of $d$.
If there is no such dot, return to the right 
(periodic boundary condition) and $d'$ is the rightmost one.

\noindent
(ii-2) Repeat (ii-1) for all the remaining unconnected dots 
in ${\bf i}$.

\noindent
(iii) The image ${\bf b} \otimes {\bf a}$ is obtained by shifting 
the $(m-l)$ unconnected dots in ${\bf j}$ to the bottom.
In the above example, 
${\bf b} \otimes {\bf a} =1110110100 \otimes 0000101110
\in B^6 \otimes B^4$. 

\begin{remark}\label{re:ny}
In the step (ii), the order of picking unconnected dots in ${\bf i}$
is not unique,
but the final result is independent of it due to \cite[Prop. 3.17]{NY}.
\end{remark}

In the above diagram (ii), the dots in the bottom tableau are 
picked  for example in the order $2,1,3,4$ from the left.
The connecting lines are called $H$-line.
We remark that NY-rule for $R: B^l \otimes B^m \rightarrow B^m \otimes B^l$ 
for $l\le m$ implies  
\begin{align}\label{le}
R({\bf i} \otimes {\bf j}) = {\bf b} \otimes {\bf a}\;\;
\Rightarrow \;\,
{\bf i} \le {\bf b}\;\,\text{and}\;\, {\bf a} \le {\bf j},
\end{align}
where 
${\bf x}\le {\bf y}$ is defined by 
${\bf y}-{\bf x} \in (\Z_{\ge 0})^L$
for ${\bf x}, {\bf y} \in \Z^L$.

The NY-rule for $R^{l,m}$ with $l\ge m$ is a dual of the above.
One starts from a dot $d$ in ${\bf j}\in B^m$ (upper tableau) 
and seeks a partner $d'$ in ${\bf i}\in B^l$ (lower tableau).
The $d'$ should be the leftmost dot among those located 
weakly right of $d$ in a periodic sense.
By construction $R^{m,l}R^{l,m}=\mathrm{id}_{B^l \otimes B^m}$ 
holds for any $l$ and $m$.
Note that $R^{l,m}$ is trivial at $l=m$ in the sense that 
$R^{l,l}({\bf i} \otimes {\bf j}) = {\bf j} \otimes {\bf i}$
in agreement with Example \ref{ex:R11} with $(q,z)=(0,1)$.

It is customary to
depict the relation $R({\bf i} \otimes {\bf j}) = {\bf b} \otimes {\bf a}$ as
\begin{equation}\label{X}
\begin{picture}(100,42)(0,-7)
\put(-8,22){${\bf b}$}\put(22,22){${\bf a}$}
\put(0,0){\vector(1,1){20}}\put(18,0){\vector(-1,1){20}}
\put(-6,-5){${\bf i}$}\put(21,-5){${\bf j}$}

\put(45,10){or}

\put(76,0){

\put(10,29){${\bf b}$}
\put(0,13){\vector(1,0){26}}
\put(13,0){\vector(0,1){26}}
\put(-7,10){${\bf i}$}\put(29,10){${\bf a}$}
\put(11,-10){${\bf j}$}
}

\end{picture}
\end{equation}
This formally looks same as (\ref{6v}).
Note however the arrows there carry $0,1$ while
those here do the 
elements from crystals $B^l, B^m$ which are $L$-tuples of $0,1$.

As the $q=0, z=z'=1$ corollary of (\ref{ybe})
the combinatorial $R$ also satisfies the Yang-Baxter equation:
\begin{equation}\label{cybe}
(R\otimes 1)(1\otimes R)(R\otimes 1) = (1\otimes R)(R\otimes 1) (1\otimes R).
\end{equation}
For example the action of the both sides on 
$0100 \otimes 0011 \otimes 1101 \in B^1 \otimes B^2 \otimes B^3$ leads to 
\begin{equation}
\begin{picture}(200,90)(0,0)

\put(0,75){\put(0,0){0111} \put(30,0){1100} \put(60,0){0001}}

\put(0,50){
\put(13.5,9){\vector(3,2){20}}\put(33.5,9){\vector(-3,2){20}}\put(70,9){\vector(0,1){12}}
}

\put(0,50){\put(0,0){0110} \put(30,0){1101} \put(60,0){0001}}

\put(0,25){
\put(10,9){\vector(0,1){12}}\put(43.5,9){\vector(3,2){20}}\put(63.5,9){\vector(-3,2){20}}
}

\put(0,25){\put(0,0){0110} \put(30,0){0001} \put(60,0){1101}}

\put(13.5,9){\vector(3,2){20}}\put(33.5,9){\vector(-3,2){20}}\put(70,9){\vector(0,1){12}}

\put(0,0){0100} \put(30,0){0011}\put(60,0){1101}

\put(95,40){$=$}

\put(120,0){
\put(0,75){\put(0,0){0111} \put(30,0){1100} \put(60,0){0001}}

\put(0,50){
\put(10,9){\vector(0,1){12}}\put(43.5,9){\vector(3,2){20}}\put(63.5,9){\vector(-3,2){20}}
}

\put(0,50){\put(0,0){0111} \put(30,0){1000} \put(60,0){0101}}

\put(0,25){
\put(13.5,9){\vector(3,2){20}}\put(33.5,9){\vector(-3,2){20}}\put(70,9){\vector(0,1){12}}
}

\put(0,25){\put(0,0){0100} \put(30,0){1011} \put(60,0){0101}}

\put(10,9){\vector(0,1){12}}\put(43.5,9){\vector(3,2){20}}\put(63.5,9){\vector(-3,2){20}}

\put(0,0){0100} \put(30,0){0011}\put(60,0){1101}
}

\end{picture}
\end{equation}
Here $=$ means that starting from the same bottom line
one ends up with the same top line despite the different order of
applications of the combinatorial $R$'s.
Combinatorial $R$'s constitute the most decent and systematic examples of 
set theoretical solutions of the Yang-Baxter equation \cite{D92} 
connected to the crystal base of quantum groups, which have 
numerous applications \cite{HKOTT,HK,IKT,KMN1,KMN2,NY}.

\subsection{\mathversion{bold}Matrix product representation of combinatorial $R$}
\label{ss:mpacR}

Setting $q = 0$ in (\ref{RL}) leads to a
matrix product representation of the combinatorial $R$.
Let us write it out in terms of the 3D $L$ operator and $q$-oscillator at $q=0$.
We will be exclusively concerned with $R^{l,m}$ with $l<m$.

First we define the $q=0$-oscillator 
${\mathscr A}_0$ to be the algebra generated by 
${\bf a}^+, {\bf a}^-, {\bf k}$ satisfying the relations\footnote{
Although the notations ${\bf a}^{\pm}, {\bf k}, F$, etc 
are retained, they are to be distinguished 
from those for ${\mathscr A}_q$.}
\begin{align}
{\bf k}^2 = {\bf k},
\quad{\bf k}\,{\bf a}^+ = 0,
\quad {\bf a}^-{\bf k}=0,
\quad{\bf a}^-{\bf a}^+ = 1,
\quad {\bf a}^+{\bf a}^- = 1-{\bf k}.
\label{ac02}
\end{align}
They may be identified with the 
operators acting on the Fock space 
$F = \bigoplus_{m \ge 0}\C |m\rangle$ as
\begin{align}
{\bf a}^+|m\rangle = |m+1\rangle,\quad
{\bf a}^-|m\rangle = (1-\delta_{m,0})|m-1\rangle,\quad
{\bf k}|m\rangle = \delta_{m,0}|m\rangle.
\label{ac01}
\end{align}
As a $\mathbb{C}$-vector space, 
the ${\mathscr A}_0$ has a Poincar\'e-Birkhoff-Witt type basis:
\begin{align}\label{pbw}
1,\quad ({\bf a}^+)^r,\quad 
({\bf a}^-)^r,\quad
({\bf a}^+)^s{\bf k}\,({\bf a}^-)^t,
\end{align}
where $r \ge 1$ and $s,t \ge 0$.
We introduce the 3D $L$ operator at $q=0$ denoted by 
$L = (L^{a,b}_{i,j})$ with indices from  
$\{0,1\}$.
Explicitly they are zero except the following:
\begin{align}\label{lak0}
L_{0, 0}^{0,0}&= L_{1,1}^{1,1} = 1,\;\;
L_{0,1}^{0,1} = {\bf k},\;\;
L_{1,0}^{0,1} = {\bf a}^+,\;\;
L^{1,0}_{0,1} = {\bf a}^-.
\end{align}
\begin{equation}\label{5v}
\begin{picture}(200,65)(-30,-36)

\put(-60,0){
\put(-52,-3){$L^{a,b}_{i,j}\;=$}
\put(-10,0){\vector(1,0){20}}
\put(0,-10){\vector(0,1){20}}
\put(-17,-3.5){$i$}\put(12.5,-3){$a$}\put(-2.4,13){$b$}\put(-2.3,-19){$j$}
}

\put(-10,0){\vector(1,0){20}}
\put(0,-10){\vector(0,1){20}}
\put(-17,-3.5){0}\put(12.5,-3.5){0}\put(-2.4,13){0}\put(-2.3,-19){0}
\put(-2,-34){1}

\put(50,0){
\put(-10,0){\vector(1,0){20}}
\put(0,-10){\vector(0,1){20}}
\put(-17,-3.5){1}\put(12.5,-3.5){1}\put(-2.4,13){1}\put(-2.3,-19){1}
\put(-2,-34){1}}

\put(100,0){
\put(-10,0){\vector(1,0){20}}
\put(0,-10){\vector(0,1){20}}
\put(-17,-3.5){1}\put(12.5,-3.5){0}\put(-2.4,13){1}\put(-2.3,-19){0}
\put(-2.5,-34){${\bf a}^+$}}

\put(150,0){
\put(-10,0){\vector(1,0){20}}
\put(0,-10){\vector(0,1){20}}
\put(-17,-3.5){0}\put(12.5,-3.5){1}\put(-2.4,13){0}\put(-2.3,-19){1}
\put(-2.5,-34){${\bf a}^-$}}

\put(200,0){
\put(-10,0){\vector(1,0){20}}
\put(0,-10){\vector(0,1){20}}
\put(-17,-3.5){0}\put(12.5,-3.5){0}\put(-2.4,13){1}\put(-2.3,-19){1}
\put(-2.5,-34){${\bf k}$}}

\end{picture}
\end{equation}
Note that compared with ${\mathscr L}$ (\ref{lak}),
the ``vertex" $L^{1,0}_{1,0}$ 
(the rightmost one in (\ref{6v})) is missing due to $q=0$.
As the result $L$ is regarded as a $q=0$-oscillator valued 
{\em five}-vertex model 
whose Boltzmann weights belong to ${\mathscr A}_0$.

Let $F^\ast = \bigoplus_{m\ge 0}\C \langle m |$ be the dual of $F$
defined by $\langle m | m'\rangle = \delta_{m,m'}$.
Let ${\mathscr A}^{\mathrm{fin}}_0 \subset {\mathscr A}_0$ 
be the vector subspace spanned by (\ref{pbw}) except 1.
Then $\mathrm{Tr}(X) := \sum_{m\ge 0}\langle m|X|m\rangle$
is finite for any $X \in {\mathscr A}^{\mathrm{fin}}_0$.
From Theorem \ref{RL} and these definitions we have
\begin{corollary}\label{co:cR}
Elements of the combinatorial 
$R=R^{l,m}: B^l \otimes B^m \rightarrow B^m \otimes B^l$ 
(\ref{sact0}) with $l<m$ are expressed in the matrix product form
\begin{align}\label{tRL}
R^{{\bf a},{\bf b}}_{{\bf i},{\bf j}}
= \mathrm{Tr}\bigl(
L^{a_1,b_1}_{i_1, j_1} \cdots L^{a_L,b_L}_{i_L, j_L} \bigr).
\end{align}
\end{corollary}

For example
$R^{100,011}_{010,101}=\mathrm{Tr}
\bigl(L^{1,0}_{0,1}L^{0,1}_{1,0}L^{0,1}_{0,1}\bigr)
=\mathrm{Tr}({\bf a}^-{\bf a}^+{\bf k}) = 1$ in agreement with 
Example \ref{cR12}.
From (\ref{lak0}) the number of ${\bf k}$'s contained in the 
above product is given by 
$|{\bf b}|-|{\bf i}| = |{\bf j}|-|{\bf a}| = m-l>0$.
Therefore 
$L^{a_1,b_1}_{i_1, j_1} \cdots L^{a_L,b_L}_{i_L, j_L} \in 
{\mathscr A}^{\mathrm{fin}}_0$ is guaranteed and the trace is convergent.
The absence of the vertex $L^{1,0}_{1,0}$ in (\ref{lak0}) 
reflects the fact that there is no dot going {\em up} in the two row diagram
in the NY-rule of $R^{l,m}$ with $l<m$.
In the product (\ref{tRL}), one can interpret that
${\bf a}^+$ creates (emits) an $H$-line from a dot in the lower tableau
and ${\bf a}^-$ annihilates (absorbs) an $H$-line into a dot in the upper tableau.
The state $|m\rangle \in F$ corresponds to the ``segment" where
there are $m$ $H$-lines.

The $L$ operator $L^{a,b}_{i,j}$ (\ref{lak0})
and the matrix product form of the combinatorial $R$ (\ref{tRL})
may be depicted in the 3D picture as follows:
\begin{equation}\label{RLpic}
\begin{picture}(220,82)(-210,-40)

\put(-210,5){
\put(-75,-7){$L^{a,b}_{i,j}\;=$}

\rotatebox{20}{
{\linethickness{0.2mm}
\put(3,1){\color{blue}\vector(-1,0){40}}}}

\put(-15,-18){\vector(0,1){32}}
\put(-30,0){\vector(3,-1){33}}

\put(-36,0){$\scriptstyle{i}$}
\put(-16,17){$\scriptstyle{b}$}
\put(-16,-26){$\scriptstyle{j}$}
\put(5,-14){$\scriptstyle{a}$}
}

%%%%%%%%%%%%% R %%%%%%%%%%%%%%%%%%%

\put(-10,4){
\put(3,1){\color{blue}\vector(-3,-1){73}}
\put(-142,-6){$R^{{\bf a},{\bf b}}_{{\bf i},{\bf j}} \;= $}

\put(-96,-26){$\mathrm{Tr}\Bigl($}
\put(68,18){$\Bigr)$}

\put(-48,-32){\vector(0,1){34}}
\put(-65,-11){\vector(3,-1){35}}
\put(-51,6){$\scriptstyle{b_1}$}
\put(-74,-10){$\scriptstyle{i_1}$}
\put(-30,-29){$\scriptstyle{a_1}$}
\put(-50,-39){$\scriptstyle{j_1}$}

\put(33,11){
\put(-48,-30){\vector(0,1){32}}
\put(-62,-12){\vector(3,-1){31}}}

\put(-38,0){$\scriptstyle{i_2}$}
\put(-18,17){$\scriptstyle{b_2}$}
\put(-17,-25){$\scriptstyle{j_2}$}
\put(4,-16){$\scriptstyle{a_2}$}

\multiput(5.1,1.7)(3,1){7}{\color{blue}.} 
\put(6,2){
\put(21,7){\color{blue}\line(3,1){30}
}

\put(83,27){
\put(-48,-27){\vector(0,1){26}}
\put(-59,-12){\vector(3,-1){25}}}
\put(15,16){$\scriptstyle{i_L}$}
\put(51,3){$\scriptstyle{a_L}$}
\put(33,29){$\scriptstyle{b_L}$}
\put(34,-7){$\scriptstyle{j_L}$}}

}

\end{picture}
\end{equation}
Here the black arrows assigned with the indices carry $0$ or $1$.
The blue ones carry the Fock space $F$
over which the trace is taken.

\section{$n$-species TASEP and its steady state}\label{sec:tasep}

\subsection{\mathversion{bold}$n$-TASEP}\label{ss:model}
Consider the periodic 1D lattice with $L$ sites which will be denoted by 
$\Z_L$.
Each site $i \in \Z_L$ is assigned with a physical variable (local state) 
$\sigma_i \in \{0,1,\ldots, n\}$.
It is interpreted as the species of the particle occupying it or 
$0$ indicating the absence of particles. 
We assume $1 \le n < L$ throughout.
Consider a stochastic model on $\Z_L$ 
such that neighboring pairs of local states
$(\sigma, \sigma') = (\sigma_i, \sigma_{i+1})$ 
are interchanged as $\sigma\, \sigma' \rightarrow \sigma'\, \sigma$
if $\sigma>\sigma'$ with the uniform transition rate.
The whole space of states is given by
\begin{align}\label{W}
(\C^{n+1})^{\otimes L} \simeq 
\bigoplus_{(\sigma_1,\ldots, \sigma_L) \in\{0,\ldots, n\}^L} 
\C|\sigma_1,\ldots, \sigma_L\kt,
\end{align}
where we suppose that 
the ket $|\cdots \rangle$ here can safely be distinguished from 
$|{\bf b}\rangle \in V^l$ for ${\bf b} \in B^l$ in Section \ref{ss:uq}
and also from $|m\rangle \in F$ for $m \in \Z_{\ge 0}$ in (\ref{ac1}) by the context.
Let $P(\sigma_1,\ldots, \sigma_L; t)$ be the probability of finding 
the configuration $(\sigma_1,\ldots, \sigma_L)$ at time $t$, and set 
\begin{align}
|P(t)\kt
= \sum_{(\sigma_1,\ldots, \sigma_L) \in\{0,\ldots, n\}^L}
P(\sigma_1,\ldots, \sigma_L; t)|\sigma_1,\ldots, \sigma_L\kt.
\end{align}
By $n$-species TASEP, or $n$-TASEP for short, we mean the stochastic system 
governed by the continuous-time master equation
\begin{align}
\frac{d}{dt}|P(t)\kt
= H |P(t)\kt,
\end{align}
where the Markov matrix 
(also called ``Hamiltonian" by abuse of terminology
despite it is not Hermitian in general)
has the form
\begin{align}\label{H}
H = \sum_{i \in \Z_L} h_{i,i+1},\qquad
h |\sigma, \sigma'\kt = \begin{cases}
|\sigma', \sigma\kt-|\sigma, \sigma'\kt & 
\;(\sigma>\sigma'),\\
0 & \; (\sigma\le \sigma').
\end{cases}
\end{align}
Here $h_{i,i+1}$ is the local Hamiltonian that  
acts as $h$ on the $i$th and the $(i+1)$th components and 
as the identity elsewhere.
As $H$ preserves the particle content, 
it acts on each {\em sector}
labeled with the 
{\em multiplicity} ${\bf m}=(m_0,\ldots, m_n) \in (\Z_{\ge 0})^{n+1}$
of the particles\footnote{$V({\bf m})$ here should not be confused with 
the antisymmetric tensor representation $V^m$  from (\ref{VBdef}).}:
\begin{align}\label{VP}
V({\bf m}) =\!\!
\sum_{{\boldsymbol \sigma} \in 
S({\bf m})}\! \C|{\boldsymbol \sigma}\kt, \quad
S({\bf m}) = 
\{{\boldsymbol \sigma}=(\sigma_1,\ldots, \sigma_L) \in \{0,\ldots, n\}^L\;|\;
\sum_{j=1}^L\delta_{k,\sigma_j}=m_k,\forall k\}. 
\end{align}
It has the dimension $\dim V({\bf m}) = \frac{L!}{m_0!\cdots m_n!}$.
The space of states is decomposed as 
$(\C^{n+1})^{\otimes L} = \bigoplus_{{\bf m}=(m_0,\ldots, m_n)}V({\bf m})$,
where the sum ranges over $m_i \in \Z_{\ge 0}$ such that 
$m_0+ \cdots + m_n = L$.
A sector $V(m_0,\ldots, m_n)$ such that $m_i \ge 1$ for all $0 \le i \le n$
is called {\em basic}.
Non-basic sectors are equivalent to a basic sector for $n'$-TASEP with some 
$n'<n$ by a suitable relabeling of species.
Thus we shall exclusively deal with basic sectors in this paper
(hence $n<L$ as mentioned before).
For various symmetry and spectral property of $H$, see \cite{AKSS}.
In each sector $V({\bf m})$ there is a unique state 
$|{\bar P}({\bf m})\kt$ up to a normalization, called the {\em steady state},  
satisfying $H|{\bar P}({\bf m})\kt = 0$.

\begin{example}\label{ex:pbar}
Up to an overall normalization, the steady states in 
$V(1,1,1), V(2,1,1)$ and $V(1,2,1)$ for 
$2$-TASEP are given by
\begin{align*} 
|{\bar P}(1,1,1)\kt& = 2 |012\kt + |021\kt 
+ |102\kt + 2 |120\kt + 
 2 |201\kt + |210\kt,\\
|{\bar P}(2,1,1)\kt& = 3 |0012\kt + |0021\kt + 2 |0102\kt
+ 3 |0120\kt + 2 |0201\kt + |0210\kt \\
&+ |1002\kt + 2 |1020\kt + 3 |1200\kt + 
 3 |2001\kt + 2 |2010\kt + |2100\kt,\\
|{\bar P}(1,2,1)\kt& = 2 |0112\kt + |0121\kt 
+ |0211\kt + 
 |1012\kt + |1021\kt + |1102\kt\\ 
 &+ 2 |1120\kt + 2 |1201\kt + |1210\kt + 
 2 |2011\kt + |2101\kt + |2110\kt.
 \end{align*}
 Similarly the steady state in $V(1,1,1,1)$ for $3$-TASEP is given by
 \begin{align*} 
 |{\bar P}(1,1,1,1)\kt& = 9 |0123\kt + 3 |0132\kt + 3 |0213\kt + 
 3 |0231\kt + 5 |0312\kt + |0321\kt + 
 3 |1023\kt + |1032\kt\\ 
 &+ 5 |1203\kt + 
 9 |1230\kt + 3 |1302\kt + 3 |1320\kt + 
 3 |2013\kt + 5 |2031\kt + |2103\kt + 
 3 |2130\kt\\ 
 &+ 9 |2301\kt + 3 |2310\kt + 
 9 |3012\kt + 3 |3021\kt + 3 |3102\kt + 
 5 |3120\kt + 3 |3201\kt + |3210\kt.
 \end{align*}
 One can observe the symmetry under the $\Z_L$ cyclic shifts
 which holds in general.
 \end{example}
 
A combinatorial algorithm to construct the steady state 
$|{\bar P}({\bf m})\kt \in V({\bf m})$ for general 
basic sectors $V({\bf m})$ of the $n$-TASEP on $\Z_L$ 
was obtained by Ferrari-Martin \cite{FM}.
In the rest of this section we show that it is most naturally  
formulated in terms of the combinatorial $R$ in the previous section.

\subsection{Multiline process}\label{ss:mlp}
To a basic sector $V({\bf m})$ of the $n$-TASEP with the multiplicity
${\bf m}= (m_0,\ldots, m_n)$, we associate another system
called {\em multiline process}. 
Define the integers $l_0,\ldots, l_{n+1}$ by
\begin{align}\label{lm}
l_i = m_{n-i+1}+\cdots + m_{n-1} + m_n\quad (0 \le i \le n+1).
\end{align}
They satisfy $0=l_0  < l_1 < \cdots < l_n < l_{n+1} = L$
due to $m_i \ge 1$ for all $i$.
Given $L$, the data $(m_i)$ and $(l_i)$ 
can be transformed to each other uniquely.
Introduce the sets
\begin{align}
{\mathcal B}({\bf m}) &=B^{l_1} \otimes \cdots \otimes B^{l_n}
= \{{\bf b}_1 \otimes \cdots \otimes {\bf b}_n\;|\; {\bf b}_j \in B^{l_j}\},
\label{Bd}\\
{\mathcal B}_+({\bf m})&=\{
{\bf b}_1 \otimes \cdots \otimes {\bf b}_n \in {\mathcal B}({\bf m})\;|\; 
{\bf b}_1 \le \cdots \le {\bf b}_n\} \subset {\mathcal B}({\bf m}),\label{Bp}
\end{align}
where $\le$ is defined after (\ref{le}).
The following bijection is elementary and will be utilized later: 
\begin{align}
&\varphi: \;S({\bf m})  \rightarrow {\mathcal B}_+({\bf m}); \quad
{\boldsymbol \sigma}=(\sigma_1,\ldots, \sigma_L) \mapsto 
\varphi_1({\boldsymbol \sigma}) \otimes \cdots \otimes \varphi_n({\boldsymbol \sigma}),\label{PB}\\
&\varphi_j({\boldsymbol \sigma}) = (\theta(\sigma_1\!\ge\! n\!+\!1\!-\!j),\ldots, 
\theta(\sigma_L\!\ge\! n\!+\!1\!-\!j)) \in B^{l_j},\label{szk}\\
&\varphi^{-1}({\bf b}_1 \otimes \cdots \otimes {\bf b}_n)=
{\bf b}_1+ \cdots +{\bf b}_n,\label{pinv}
\end{align}
where $\theta$ is defined in Example \ref{ex:R11}.
Applying $\varphi^{-1}$ to (\ref{PB}) and using (\ref{pinv})  we see that 
\begin{equation}\label{tout}
{\boldsymbol \sigma} = \varphi_1({\boldsymbol \sigma}) + 
\cdots + \varphi_n({\boldsymbol \sigma})
\end{equation}
holds identically.
If (\ref{szk}) is given as 
$\varphi_j({\boldsymbol \sigma})=(\gamma_{j,1},\ldots, \gamma_{j,L})$,
it is also easy to see that
\begin{equation}\label{Lin} 
\sigma_k=i \;\; \Rightarrow\;\;
(\gamma_{1,k},\ldots, \gamma_{n,k})
= (\overbrace{0,\ldots,0}^{n-i},\overbrace{1,\ldots,1}^i)
\end{equation}
for each $k \in \Z_L$.
The sum in (\ref{pinv}) as elements in $\Z^L$ is not an
operation usually performed in the crystal base theory.

\begin{example}\label{ex:PB}
Let $n=3$ and take ${\boldsymbol \sigma}=(3,0,1,2,3,0,1) \in S({\bf m})$
with ${\bf m} = (2,2,1,2)$.
The image $\varphi({\boldsymbol \sigma})
= \varphi_1({\boldsymbol \sigma})\otimes\varphi_2({\boldsymbol \sigma})\otimes\varphi_3({\boldsymbol \sigma}) \in 
{\mathcal B}_+({\bf m}) \subset  B^2\otimes B^3 \otimes B^5$
is given by
\begin{table}[h]
\begin{tabular}{c|c|c}
${\boldsymbol \sigma}$ & 3 0 1 2 3 0 1 & $\sigma_k$\\
\hline
$\varphi_3({\boldsymbol \sigma})$ & 1 0 1 1 1 0 1 & $\gamma_{3,k}$ \\

$\varphi_2({\boldsymbol \sigma})$ & 1 0 0 1 1 0 0 & $\gamma_{2,k}$\\

$\varphi_1({\boldsymbol \sigma})$ & 1 0 0 0 1 0 0 & $\gamma_{1,k}$
\end{tabular}
\end{table}

One can check 
$\varphi_1({\boldsymbol \sigma})\le \varphi_2({\boldsymbol \sigma}) \le \varphi_3({\boldsymbol \sigma})$ 
(\ref{Bp}), 
${\boldsymbol \sigma} = \varphi_1({\boldsymbol \sigma})+
\varphi_2({\boldsymbol \sigma})+\varphi_3({\boldsymbol \sigma})$ (\ref{tout}) and 
(\ref{Lin}).
\end{example}

The multiline process \cite{FM} is a stochastic process on the set 
${\mathcal B}({\bf m})$.
To describe its dynamics we first introduce a deterministic map
\begin{align}\label{T}
T: \;&  {\mathcal B}({\bf m}) \otimes \Z_L 
\rightarrow \Z_L \otimes {\mathcal B}({\bf m}); \qquad
{\bf b}_1 \otimes \cdots \otimes {\bf b}_n \otimes k \mapsto
k' \otimes {\bf b}'_1 \otimes \cdots \otimes {\bf b}'_n.
\end{align}
It is defined by sending $\Z_L$ from the right to the left through 
$B^{l_1} \otimes \cdots \otimes B^{l_n}$ 
by successively applying the following pairwise exchange rule:
\begin{equation}\label{LL}
\begin{split}
&B^l \otimes \Z_L \rightarrow \Z_L \otimes B^l;
\qquad (x_1,\ldots, x_L) \otimes k \mapsto
k' \otimes (x'_1,\ldots, x'_L),\\
&k' = k+x_k-1,\qquad x'_i 
= \begin{cases} 
\min(x_k,x_{k+1}) & i=k,\\
\max(x_k,x_{k+1})& i=k+1,\\
x_i & \text{otherwise}.
\end{cases}
\end{split}
\end{equation}
The map $T$ in  (\ref{T}) induces the ``time evolutions" 
on ${\mathcal B}({\bf m})$ by setting
\begin{align}\label{Tk}
T_k :\; {\mathcal B}({\bf m}) \rightarrow {\mathcal B}({\bf m}); \qquad
T_k({\bf b}_1 \otimes \cdots \otimes {\bf b}_n) = 
{\bf b}'_1 \otimes \cdots \otimes {\bf b}'_n\quad (k \in \Z_L),
\end{align}
where the right hand side is the one appearing in (\ref{T}).

\begin{example}\label{ex:tk}
Take ${\bf s} ={\bf b}_1 \otimes {\bf b}_2 \otimes {\bf b}_3 = 
000010 \otimes 001010 \otimes 001011 \in B^1 \otimes B^2 \otimes B^3$.
Then $T_2({\bf s}) = {\bf s}$ and the other cases are given by
\[
\begin{picture}(300,60)(42,-13)

\put(2.5,35){0 \,0\, 2\, 0\,  3\, 1}
\put(-16,24){${\bf b}_3$}
\put(-16,13){${\bf b}_2$}
\put(-16,2){${\bf b}_1$}
\multiput(0,0)(10,0){7}{\put(0,0){\line(0,1){30}}}
\multiput(0,0)(0,10){4}{\put(0,0){\line(1,0){60}}}
\put(22.5,23){$\bullet$}\put(42.5,23){$\bullet$}\put(52.5,23){$\bullet$}
\put(22.5,13){$\bullet$}\put(42.5,13){$\bullet$}
\put(42.5,3){$\bullet$}
\put(28,-13){${\bf s}$}

\put(67,0){
\put(2.5,35){0 \,0\, 2\, 0\,  3\, 1}
\multiput(0,0)(10,0){7}{\put(0,0){\line(0,1){30}}}
\multiput(0,0)(0,10){4}{\put(0,0){\line(1,0){60}}}
\put(22.5,23){$\bullet$}\put(42.5,23){$\bullet$}\put(52.5,23){$\bullet$}
\put(22.5,13){$\bullet$}\put(42.5,13){$\bullet$}
\put(52.5,3){$\bullet$}
\put(20,-13){$T_1({\bf s})$}
}

\put(134,0){
\put(2.5,35){0 \,0\, 0\, 2\,  3\, 1}
\multiput(0,0)(10,0){7}{\put(0,0){\line(0,1){30}}}
\multiput(0,0)(0,10){4}{\put(0,0){\line(1,0){60}}}
\put(32.5,23){$\bullet$}\put(42.5,23){$\bullet$}\put(52.5,23){$\bullet$}
\put(32.5,13){$\bullet$}\put(42.5,13){$\bullet$}
\put(42.5,3){$\bullet$}
\put(20,-13){$T_3({\bf s})$}
}

\put(201,0){
\put(2.5,35){0 \,0\, 2\, 0\,  3\, 1}
\multiput(0,0)(10,0){7}{\put(0,0){\line(0,1){30}}}
\multiput(0,0)(0,10){4}{\put(0,0){\line(1,0){60}}}
\put(22.5,23){$\bullet$}\put(42.5,23){$\bullet$}\put(52.5,23){$\bullet$}
\put(32.5,13){$\bullet$}\put(42.5,13){$\bullet$}
\put(42.5,3){$\bullet$}
\put(20,-13){$T_4({\bf s})$}
}

\put(269,0){
\put(2.5,35){0 \,0\, 2\, 0\,  1\, 3}
\multiput(0,0)(10,0){7}{\put(0,0){\line(0,1){30}}}
\multiput(0,0)(0,10){4}{\put(0,0){\line(1,0){60}}}
\put(22.5,23){$\bullet$}\put(42.5,23){$\bullet$}\put(52.5,23){$\bullet$}
\put(22.5,13){$\bullet$}\put(52.5,13){$\bullet$}
\put(52.5,3){$\bullet$}
\put(20,-13){$T_5({\bf s})$}
}

\put(336,0){
\put(2.5,35){1 \,0\, 2\, 0\,  3\, 0}
\multiput(0,0)(10,0){7}{\put(0,0){\line(0,1){30}}}
\multiput(0,0)(0,10){4}{\put(0,0){\line(1,0){60}}}
\put(2.5,23){$\bullet$}\put(22.5,23){$\bullet$}\put(42.5,23){$\bullet$}
\put(22.5,13){$\bullet$}\put(42.5,13){$\bullet$}
\put(52.5,3){$\bullet$}
\put(20,-13){$T_6({\bf s})$}
}

\end{picture}
\]
Here the same tableaux as for the NY-rule are used.
As an example, $T_3({\bf s})$ has been obtained
by applying  (\ref{LL}) successively as 
\begin{align*}
{\bf s} \otimes 3 =&\;000010 \otimes 001010 \otimes 001011 
\otimes 3\mapsto
000010 \otimes 001010 \otimes 3 \otimes 000111\\
\mapsto &\;
000010 \otimes 3 \otimes 000110 \otimes 000111 \mapsto
2 \otimes 000010\otimes 000110 \otimes 000111 = 2 \otimes T_3({\bf s}).
\end{align*}
The sequences above the tableaux will be explained after (\ref{ypi}).
\end{example}

Now we can define a stochastic dynamics on 
${\mathcal B}({\bf m})$ by declaring that 
each element of it undergoes the $L$ kinds of time evolutions 
$T_1, \ldots, T_L$ with an equal probability.
The resulting system is called multiline process \cite{FM}.
We have not found a crystal theoretic interpretation of $T_k$.
See Section \ref{ss:gd} for a further remark.

\subsection{\mathversion{bold}Projection from multiline process to $n$-TASEP}

Fix $L, n$ and ${\bf m}=(m_i)$ and $(l_i)$ as explained around (\ref{lm}).
Recall that the sector $V({\bf m})$ of the $n$-TASEP is defined in (\ref{VP}) 
and the set of states of multiline process is 
${\mathcal B}({\bf m})=B^{l_1} \otimes \cdots \otimes B^{l_n}$ in (\ref{Bd}).
We introduce a map
\begin{align}
\pi_j: {\mathcal B}({\bf m}) \rightarrow B^{l_j};\quad
{\bf s}={\bf b}_1 \otimes \cdots \otimes {\bf b}_n \mapsto \pi_j({\bf s})
\qquad(1 \le j \le n)
\end{align}
by a composition of the combinatorial $R$.
It is most easily grasped by the diagram 
\begin{equation}\label{pijdef}
\begin{picture}(100,38)(10,-13)
\put(-13,7){${\bf b}_j$}
\put(0,10){\line(1,0){42}}
\put(47,7.5){$\cdots$}
\put(65,10){\vector(1,0){25}}

\put(10,-2){\vector(0,1){24}} \put(5,-12.5){${\bf b}_{j+1}$}
\put(32,-2){\vector(0,1){24}} \put(30,-13){${\bf b}_{j+2}$}
\put(77,-2){\vector(0,1){24}} \put(75,-13){${\bf b}_{n}$}

\put(94,7){$\pi_j({\bf s})$,}
\end{picture}
\end{equation}
which is a composition of (\ref{X}).
Namely one lets the component $B^{l_j}$ penetrate through its right 
neighbor $B^{l_{j+1}} \otimes \cdots \otimes B^{l_n}$ 
by successively applying the combinatorial $R=R^{l_j, l_r}$
for $r=j+1, \cdots, n$.
Note that $\pi_j({\bf b}_1 \otimes \cdots \otimes {\bf b}_n)$
actually depends only on 
${\bf b}_j \otimes \cdots \otimes {\bf b}_n$ and in particular  
$\pi_n({\bf b}_1 \otimes \cdots \otimes {\bf b}_n)={\bf b}_n$.

From $l_1<\cdots < l_n$ (see around (\ref{lm})) 
and (\ref{le}), we have an increasing sequence 
$\pi_1({\bf s}) \le 
\pi_2({\bf s}) \le
\cdots \le 
\pi_n({\bf s})$ of elements of the 
crystals $B^{l_1}, B^{l_2}, \ldots, B^{l_n}$.
In fact one can check
\begin{align}\label{ypi}
\pi_1({\bf s})\otimes  \cdots \otimes \pi_n({\bf s}) \in 
{\mathcal B}_+({\bf m})
\end{align}
for any ${\bf s} \in {\mathcal B}({\bf m})$.
Sending this by $\varphi^{-1}$ and applying (\ref{pinv}), 
we are able to define
a surjective map $\pi$ from the states 
${\mathcal B}({\bf m})$ (\ref{Bd}) of the 
multiline process to the states $S({\bf m})$ (\ref{VP})  in the 
$n$-TASEP by
\begin{align}\label{tpdef}
\pi : {\mathcal B}({\bf m}) \rightarrow 
S({\bf m});\quad
\pi({\bf s}) = \pi_1({\bf s})+  \cdots + \pi_n({\bf s}).
\end{align}
The sequences above the tableaux in Example \ref{ex:tk} 
show the images in $S(3,1,1,1)$ under $\pi$.
By comparing the NY-rule and the diagram (\ref{pijdef}) with 
Fig.2 and Fig 3 in \cite{FM} and noting Remark \ref{re:ny},  
it can be seen that the map $\pi$ 
coincides with the Ferrari-Martin algorithm 
$V^{(L)}: {\mathcal X} \rightarrow {\mathcal V}$ in  \cite{FM}
upon conventional adjustment.
This is our first main observation in this paper.

As a consequence of this identification, 
Theorem 4.1 and Theorem 2.2 in \cite{FM} lead to  
Proposition \ref{pr:time} and Proposition \ref{pr:fm} given below.

\begin{proposition}\label{pr:time}
Let $\tau_i$ 
be the map on $S({\bf m})$ that changes the 
$i$th and $(i\!+\!1)$th components 
$(\sigma_i, \sigma_{i+1})$ of ${\boldsymbol \sigma}  \in S({\bf m})$ into 
$(\min(\sigma_i, \sigma_{i+1}),\max(\sigma_i, \sigma_{i+1}))$.
Then the following diagram is commutative:
\begin{equation*}
\begin{CD}
{\mathcal B}({\bf m}) @> {T_i} >> {\mathcal B}({\bf m}) \\
@V{\pi}VV  @V{\pi}VV\\
S({\bf m}) @> \tau_i>> S({\bf m}).
\end{CD}
\end{equation*}
\end{proposition}
One can easily check the commutativity in Example \ref{ex:tk}.
See Section \ref{ss:gd} for a crystal theoretical 
generalization of $T_i$ and $\tau_i$.

\begin{proposition}\label{pr:fm}
Up to a normalization, the steady state in the sector $V({\bf m})$ is given by 
\begin{align}
&|\bar{P}({\bf m})\kt 
= \sum_{{\bf s} \in {\mathcal B}({\bf m})}
|\pi({\bf s})\kt =
\sum_{{\boldsymbol \sigma} = (\sigma_1,\ldots, \sigma_L) \in S({\bf m})} 
\mathbb{P}({\boldsymbol \sigma})|{\boldsymbol \sigma}\kt,
\nonumber\\
&\mathbb{P}({\boldsymbol \sigma}) = \#\{{\bf s} 
\in {\mathcal B}({\bf m})\;|\;
\pi({\bf s})= {\boldsymbol \sigma}\}.
\label{pbar}
\end{align}
\end{proposition}

The last condition is equivalent to 
$\varphi^{-1}(\pi_1({\bf s}) \otimes \cdots \otimes \pi_n({\bf s}) )
= {\boldsymbol \sigma}$ due to (\ref{pinv}) and (\ref{tpdef}).
By taking $\varphi$ further and using (\ref{PB}) 
we find 
\begin{align}\label{pbar2}
\mathbb{P}({\boldsymbol \sigma} )=
\#\{ {\bf s} \in {\mathcal B}({\bf m})\;|\;
\pi_k({\bf s}) = \varphi_k({\boldsymbol \sigma})\;\;(1 \le k \le n)\}.
\end{align}
In the next section we invoke the results in Section \ref{ss:mpacR} 
to derive a new matrix product formula for the (unnormalized)
steady state probability $\mathbb{P}({\boldsymbol \sigma})$ of the 
configuration ${\boldsymbol \sigma}$.

\section{Matrix product formula}\label{sec:mpa}

\subsection{\mathversion{bold}Diagram for $\pi_i$: Combinatorial 
corner transfer matrix}\label{ss:ctm}

There is a {\em single} diagram that represents 
all of $\pi_1({\bf s}), \ldots, \pi_n({\bf s})$ in (\ref{pijdef})
simultaneously.
We illustrate it along $n=3$ case.
Given ${\bf s} = {\bf b}_1 \otimes{\bf b}_2 \otimes{\bf b}_3 
\in {\mathcal B}({\bf m}) = B^{l_1} \otimes B^{l_2} \otimes B^{l_3}$,  
the diagrams (\ref{pijdef}) for $\pi_1$ and $\pi_2$ are
\begin{equation*}
\begin{picture}(250,40)(-85,-15)

\put(-45,0){
\put(-13.5,7){${\bf b}_1$}
\put(0,10){\vector(1,0){42}}
\put(10,-2){\vector(0,1){24}} \put(5,-12.5){${\bf b}_{2}$}
\put(32,-2){\vector(0,1){24}} \put(30,-13){${\bf b}_{3}$}
\put(46,7){$\pi_1({\bf s})$}
}

\put(50,7){${\bf b}_2$}
\put(65,10){\vector(1,0){25}}
\put(77,-2){\vector(0,1){24}} \put(75,-13){${\bf b}_3$}

\put(94,7){$\pi_2({\bf s})$}
\end{picture}
\end{equation*}
Here each vertex stands for the combinatorial $R$. See (\ref{X}).
This defining diagram of $\pi_1({\bf s})$ can be deformed 
by means of the Yang-Baxter equation (\ref{cybe}):
\begin{equation}\label{ptanj}
\begin{picture}(250,45)(-70,-15)

\put(-40,0){

\put(-13.5,7){${\bf b}_1$}
\put(0,10){\vector(1,0){42}}
\put(46,7){$\pi_1({\bf s})$}

\put(0,-5){
\put(0,25){\put(10,-3.2){\vector(2,1){22}}\put(32,-2.4){\vector(-2,1){22}}}
\put(10,-2){\line(0,1){24}} \put(5,-12.5){${\bf b}_{2}$}
\put(32,-2){\line(0,1){24}} \put(30,-13){${\bf b}_{3}$}
}
}

\put(80,0){\put(-38,7){$=$}
\put(-13.5,7){${\bf b}_1$}
\put(0,10){\vector(1,0){42}}
\put(10,4){\vector(0,1){22}} 
\put(32,4){\vector(0,1){22}} 

\put(0,-4){
\put(10,-3.2){\line(2,1){22}}\put(32,-2.8){\line(-2,1){22}}
\put(5,-12.5){${\bf b}_{2}$}\put(30,-12.5){${\bf b}_{3}$}
}

\put(46,7){$\pi_1({\bf s})$}
}
\end{picture}
\end{equation}
Inclining the vertices appropriately in the last diagram  
we find that 
$\pi_1({\bf s}), \pi_2({\bf s}), \pi_3({\bf s})(={\bf b}_3)$
can all be included in the left diagram below:
\begin{equation}\label{ctm1}
\begin{picture}(120,88)(-15,-12)

\put(-60,10){
\put(10,58){$n=3$}
\put(0,40){\line(1,0){50}} \put(50,40){\vector(0,1){10}}
\put(0,20){\line(1,0){30}} \put(30,20){\vector(0,1){30}}
\put(0,0){\line(1,0){10}}
\put(10,0){\vector(0,1){50}}

\put(-15,38){${\bf b}_1$}
\put(-15,18){${\bf b}_2$}
\put(-15,-2){${\bf b}_3$}

\put(55,38){$\pi_1({\bf s})$}
\put(35,18){$\pi_2({\bf s})$}
\put(15,-2){$\pi_3({\bf s})$}}

\put(80,-10){
\put(20,78){$n=4$}
\put(0,60){\line(1,0){70}} \put(70,60){\vector(0,1){10}}
\put(0,40){\line(1,0){50}} \put(50,40){\vector(0,1){30}}
\put(0,20){\line(1,0){30}} \put(30,20){\vector(0,1){50}}
\put(0,0){\line(1,0){10}}
\put(10,0){\vector(0,1){70}}

\put(-15,58){${\bf b}_1$}
\put(-15,38){${\bf b}_2$}
\put(-15,18){${\bf b}_3$}
\put(-15,-2){${\bf b}_4$}

\put(75,58){$\pi_1({\bf s})$}
\put(55,38){$\pi_2({\bf s})$}
\put(35,18){$\pi_3({\bf s})$}
\put(15,-2){$\pi_4({\bf s})$}}

\end{picture}
\end{equation}
The result for $n=4$ case is also given together for which 
${\bf s}$ should be understood as 
${\bf s} = {\bf b}_1\otimes \cdots \otimes {\bf b}_4
\in {\mathcal B}({\bf m}) = B^{l_1} \otimes \cdots \otimes B^{l_4}$. 
General case is clear and similarly obtained by 
transforming the defining diagrams (\ref{pijdef}) of $\pi_j({\bf s})$ 
for $j=1,\ldots, n$ into composable forms 
with the aid of the Yang-Baxter equation as in (\ref{ptanj}).

In (\ref{ctm1}) the arrow starting from ${\bf b}_i$ 
carries the crystal $B^{l_i}$ and 
each segment corresponds to an element of it. 
The only nontrivial events are the combinatorial $R$'s 
indicated by the vertices.
The $90^\circ$ left turns along the arrows do not change the
elements.
It is consistent with the bottom line since 
$\pi_n({\bf s})={\bf b}_n$ as noted after (\ref{pijdef}).
The inclusion of such turns makes the diagram symmetric
in that there are $n$ incoming arrows horizontally and 
$n$ outgoing ones vertically.
A further significance will become manifest in \cite{KMO}. 
Since the combinatorial $R$ is a deterministic map,
every part of the diagram is fixed uniquely upon choosing the input 
${\bf s} = {\bf b}_1 \otimes \cdots \otimes{\bf b}_n 
\in {\mathcal B}({\bf m})$ from the left.

In this way we are led to the 
{\em corner transfer matrix} \cite[Chap.13]{Bax}
of the vertex model associated with 
the antisymmetric tensor representations of $U_q(\widehat{sl}_L)$.
It is at the combinatorial point $q=0$ where every vertex is crystallized to 
the combinatorial $R$.
Note that the physical space $\Z_L$ and the internal degrees of freedom 
$\{0,1,\ldots, n\}$ in the original $n$-TASEP have been interchanged;
we have a corner transfer matrix on the system of linear size $n$
whose local interaction is encoded in the 
quantum group $U_q(\widehat{sl}_L)$ at $q=0$.
We stress that such a cross-channel of the problem has been captured precisely 
by reformulating the Ferrari-Martin algorithm in terms of combinatorial $R$ 
and thereby enabling a systematic use of the Yang-Baxter equation.

\subsection{Factorization into five-vertex model}\label{ss:mpr}

By the graphical representation (\ref{ctm1}) of $\pi_k({\bf s})$, the 
stationary probability (\ref{pbar2}) for the configuration 
${\boldsymbol \sigma}=(\sigma_1,\ldots, \sigma_L)$ is expressed,  
for instance for $n=3$, as
\begin{equation}\label{pbar3}
\begin{picture}(120,65)(-50,-9)

\put(-122,25){${\displaystyle \mathbb{P}({\boldsymbol \sigma}) 
= \sum_{{\bf b}_1\otimes {\bf b}_2\otimes {\bf b}_3
\in {\mathcal B}({\bf m})}}$}
\put(0,40){\line(1,0){50}} \put(50,40){\vector(0,1){10}}
\put(0,20){\line(1,0){30}} \put(30,20){\vector(0,1){30}}
\put(0,0){\line(1,0){10}}
\put(10,0){\vector(0,1){50}}

\put(-15,38){${\bf b}_1$}
\put(-15,18){${\bf b}_2$}
\put(-15,-2){${\bf b}_3$}

\put(55,38){$\varphi_1({\boldsymbol \sigma})$}
\put(35,18){$\varphi_2({\boldsymbol \sigma})$}
\put(15,-2){$\varphi_3({\boldsymbol \sigma})$}

\end{picture}
\end{equation}
Here each $\varphi_j({\boldsymbol \sigma})$ signifies
the {\em boundary condition} 
$\pi_j({\bf b}_1\otimes {\bf b}_2\otimes {\bf b}_3)
= \varphi_j({\boldsymbol \sigma})$
that the element $\pi_j({\bf b}_1\otimes {\bf b}_2\otimes {\bf b}_3) 
\in B^{l_j}$ 
making the left turn there should coincide with 
$\varphi_j({\boldsymbol \sigma})$ prescribed from 
${\boldsymbol \sigma}$ 
by the simple rule (\ref{szk}).
Each summand of (\ref{pbar3}) means 1 or 0
depending on whether such a boundary condition is 
satisfied or not for the chosen 
${\bf b}_1\otimes {\bf b}_2\otimes {\bf b}_3$.

Now we are in the position to convert this formula into 
a matrix product form by invoking Corollary \ref{co:cR}.
To each vertex of (\ref{pbar3}),  
substitute  (\ref{tRL}) or equivalently 
the graphical form (\ref{RLpic}) of the combinatorial $R$.
It amounts to putting the three ``skewers" standing upright at the vertices.
The result takes the form
\begin{align}\label{ptr}
\mathbb{P}
({\boldsymbol \sigma}) = \mathrm{Tr}\bigl(X_{\sigma_1} \cdots X_{\sigma_L}\bigr),
\end{align}
where the trace extends over $F^{\otimes 3}$ 
and the factor $X_{\sigma_k}$ corresponds to the $k$th layer of the skewers.
It has the same structure as 
(\ref{pbar3}) but now each vertex represents an $\mathscr{A}_0$-valued 
Boltzmann weight of the five-vertex model (\ref{5v}).
Accordingly each arrow carries $0$ or $1$, and the sum over 
${\bf b}_1\otimes{\bf b}_2\otimes{\bf b}_3$ should be replaced by the  
{\em configuration sum} of the five-vertex model.
What about the boundary condition on them?
It should be imposed so as to reflect  
the table of Example \ref{ex:PB}  in the diagram (\ref{pbar3}).
In this way we find
\begin{equation}\label{X3}
\begin{picture}(600,43)(-59,0)
\setlength\unitlength{0.26mm}

\put(-56,24){$X_0= \sum$}
\put(0,40){\line(1,0){50}} \put(50,40){\vector(0,1){10}}
\put(0,20){\line(1,0){30}} \put(30,20){\vector(0,1){30}}
\put(0,0){\line(1,0){10}}
\put(10,0){\vector(0,1){50}}

\put(55,36){0}
\put(35,16){0}
\put(15,-4){0}

\put(140,0){
\put(-56,24){$X_1= \sum$}
\put(0,40){\line(1,0){50}} \put(50,40){\vector(0,1){10}}
\put(0,20){\line(1,0){30}} \put(30,20){\vector(0,1){30}}
\put(0,0){\line(1,0){10}}
\put(10,0){\vector(0,1){50}}

\put(55,36){0}
\put(35,16){0}
\put(15,-4){1}}

\put(280,0){
\put(-56,24){$X_2= \sum$}
\put(0,40){\line(1,0){50}} \put(50,40){\vector(0,1){10}}
\put(0,20){\line(1,0){30}} \put(30,20){\vector(0,1){30}}
\put(0,0){\line(1,0){10}}
\put(10,0){\vector(0,1){50}}

\put(55,36){0}
\put(35,16){1}
\put(15,-4){1}}

\put(420,0){
\put(-56,24){$X_3= \sum$}
\put(0,40){\line(1,0){50}} \put(50,40){\vector(0,1){10}}
\put(0,20){\line(1,0){30}} \put(30,20){\vector(0,1){30}}
\put(0,0){\line(1,0){10}}
\put(10,0){\vector(0,1){50}}
\put(55,36){1}
\put(35,16){1}
\put(15,-4){1}}

\end{picture}
\end{equation}
The sums here extend over all the configurations of the 
five-vertex model under the specified boundary conditions.
The resulting objects take values in ${\mathscr A}_0^{\otimes 3}$.
Explicitly they read
\begin{equation*}
\begin{picture}(600,60)(-59,-20)
\setlength\unitlength{0.26mm}

\put(-40,24){$X_0= $}
\put(0,40){\line(1,0){50}} \put(50,40){\vector(0,1){10}}
\put(0,20){\line(1,0){30}} \put(30,20){\vector(0,1){30}}
\put(0,0){\line(1,0){10}}
\put(10,0){\vector(0,1){50}}
\put(-18,-20){$=1\otimes 1 \otimes 1$}

\put(90,0){\put(-20,24){$+$}
\put(10,40){\line(1,0){40}} \put(50,40){\vector(0,1){10}}
\put(0,20){\line(1,0){30}} \put(30,20){\vector(0,1){30}}
\put(0,0){\line(1,0){10}}
\put(10,0){\line(0,1){40}}
\put(0,40){\color{red}\line(1,0){10}}
\put(10,40){\color{red}\vector(0,1){10}}
\put(-22,-20){$+\;\; {\bf a}^+\otimes 1 \otimes 1$}}

\put(185,0){\put(-25,24){$+$}
\put(0,40){\line(1,0){50}} \put(50,40){\vector(0,1){10}}
\put(0,20){\color{red}{\line(1,0){10}}} \put(10,20){\line(1,0){20}}
\put(30,20){\vector(0,1){30}}
\put(0,0){\line(1,0){10}}
\put(10,0){\line(0,1){20}}\put(10,20){\color{red}\vector(0,1){30}}
\put(-22,-20){$+\;\; {\bf k}\otimes {\bf a}^+ \otimes 1$}}

\put(285,0){\put(-30,24){$+$}
\put(0,40){\line(1,0){10}} \put(10,40){\color{red}\line(1,0){20}}\put(30,40){\line(1,0){20}} 
\put(50,40){\vector(0,1){10}}\put(10,40){\vector(0,1){10}}\put(10,20){\color{red}\line(0,1){20}}
\put(0,20){\color{red}\line(1,0){10}}
\put(10,20){\line(1,0){20}} \put(30,40){\color{red}\vector(0,1){10}}
\put(30,20){\line(0,1){20}}
\put(0,0){\line(1,0){10}}
\put(10,0){\line(0,1){20}}
\put(-29,-20){$+\;\; {\bf a}^-\otimes {\bf a}^+ \otimes {\bf a}^+$}}

\put(390,0){\put(-25,24){$+$}
\put(0,20){\color{red}\line(1,0){10}}\put(10,20){\color{red}\vector(0,1){30}}
\put(0,40){\color{red}\line(1,0){30}}\put(30,40){\color{red}\vector(0,1){10}}
\put(30,40){\line(1,0){20}} \put(50,40){\vector(0,1){10}}
\put(10,20){\line(1,0){20}} \put(30,20){\line(0,1){20}}
\put(0,0){\line(1,0){10}}
\put(10,0){\line(0,1){20}}
\put(-25,-20){$+\;\; 1\otimes {\bf a}^+ \otimes {\bf a}^+$}}

\end{picture}
\end{equation*}
\begin{equation*}
\begin{picture}(600,60)(-59,-20)
\setlength\unitlength{0.26mm}

\put(-40,24){$X_1= $}
\put(0,40){\line(1,0){50}} \put(50,40){\vector(0,1){10}}
\put(0,20){\line(1,0){30}} \put(30,20){\vector(0,1){30}}
\put(0,0){\color{red}\line(1,0){10}}
\put(10,0){\color{red}\vector(0,1){50}}
\put(-18,-20){$={\bf k} \otimes {\bf k} \otimes 1$}

\put(90,0){\put(-20,24){$+$}
\put(10,40){\color{red}\line(1,0){20}} \put(30,40){\line(1,0){20}}
\put(50,40){\vector(0,1){10}}\put(30,40){\color{red}\vector(0,1){10}}
\put(0,20){\line(1,0){30}} \put(30,20){\line(0,1){20}}
\put(0,0){\color{red}\line(1,0){10}}
\put(10,0){\color{red}\line(0,1){40}}
\put(0,40){\line(1,0){10}}
\put(10,40){\vector(0,1){10}}
\put(-22,-20){$+\;\; {\bf a}^-\otimes {\bf k} \otimes {\bf a}^+$}}

\put(185,0){\put(-20,24){$+$}
\put(50,40){\vector(0,1){10}}\put(30,40){\color{red}\vector(0,1){10}}
\put(0,40){\color{red}\line(1,0){30}}\put(30,40){\line(1,0){20}}
\put(30,20){\line(0,1){20}}
\put(0,20){\line(1,0){30}}
\put(10,0){\color{red}\vector(0,1){50}}
\put(0,0){\color{red}\line(1,0){10}}
\put(-22,-20){$+\;\; 1\otimes {\bf k} \otimes {\bf a}^+$}}

\end{picture}
\end{equation*}
\begin{equation*}
\begin{picture}(600,60)(-59,-20)
\setlength\unitlength{0.26mm}

\put(-40,24){$X_2= $}
\put(0,40){\line(1,0){50}}\put(50,40){\vector(0,1){10}}
\put(0,20){\line(1,0){10}}\put(10,20){\color{red}\line(1,0){20}}
\put(30,20){\color{red}\vector(0,1){30}}\put(10,20){\vector(0,1){30}}
\put(10,0){\color{red}\line(0,1){20}}
\put(0,0){\color{red}\line(1,0){10}}
\put(-22,-20){$\;=\;1\otimes {\bf a}^- \otimes {\bf k}$}

\put(100,0){\put(-28,24){$+$}
\put(0,40){\color{red}\line(1,0){10}}\put(10,40){\line(1,0){40}}
\put(50,40){\vector(0,1){10}}
\put(0,20){\line(1,0){10}}\put(10,20){\color{red}\line(1,0){20}}
\put(30,20){\color{red}\vector(0,1){30}}\put(10,40){\color{red}\vector(0,1){10}}
\put(10,20){\line(0,1){20}}
\put(10,0){\color{red}\line(0,1){20}}
\put(0,0){\color{red}\line(1,0){10}}
\put(-32,-20){$\;+\;\;{\bf a}^+\otimes {\bf a}^- \otimes {\bf k}$}
}

\put(200,0){\put(-25,24){$+$}
\put(0,40){\line(1,0){50}} \put(50,40){\vector(0,1){10}}
\put(0,20){\color{red}\line(1,0){30}} \put(30,20){\color{red}\vector(0,1){30}}
\put(0,0){\color{red}\line(1,0){10}}
\put(10,0){\color{red}\vector(0,1){50}}
\put(-23,-20){$+\;\;\;{\bf k}\otimes 1 \otimes {\bf k}$}}

\end{picture}
\end{equation*}
\begin{equation*}
\begin{picture}(600,60)(-59,-20)
\setlength\unitlength{0.26mm}

\put(-40,24){$X_3= $}
\put(0,20){\line(1,0){10}}\put(10,20){\vector(0,1){30}}
\put(0,40){\line(1,0){30}}\put(30,40){\vector(0,1){10}}

\put(30,40){\color{red}\line(1,0){20}} \put(50,40){\color{red}\vector(0,1){10}}
\put(10,20){\color{red}\line(1,0){20}} \put(30,20){\color{red}\line(0,1){20}}
\put(0,0){\color{red}\line(1,0){10}}
\put(10,0){\color{red}\line(0,1){20}}
\put(-27,-20){$=1\otimes {\bf a}^- \otimes {\bf a}^-$}

\put(90,0){\put(-30,24){$+$}
\put(0,40){\color{red}\line(1,0){10}} \put(10,40){\line(1,0){20}}\put(30,40){\color{red}\line(1,0){20}} 
\put(50,40){\color{red}\vector(0,1){10}}\put(10,40){\color{red}\vector(0,1){10}}\put(10,20){\line(0,1){20}}
\put(0,20){\line(1,0){10}}
\put(10,20){\color{red}\line(1,0){20}} \put(30,40){\vector(0,1){10}}
\put(30,20){\color{red}\line(0,1){20}}
\put(0,0){\color{red}\line(1,0){10}}
\put(10,0){\color{red}\line(0,1){20}}
\put(-29,-20){$+\;\; {\bf a}^+\otimes {\bf a}^- \otimes {\bf a}^-$}}

\put(195,0){
\put(-30,24){$+$}
\put(0,40){\line(1,0){30}} \put(50,40){\color{red}\vector(0,1){10}}
\put(0,20){\color{red}\line(1,0){30}} \put(30,40){\vector(0,1){10}}
\put(0,0){\color{red}\line(1,0){10}}\put(30,20){\color{red}\line(0,1){20}}
\put(10,0){\color{red}\vector(0,1){50}}\put(30,40){\color{red}\line(1,0){20}}
\put(-28,-20){$+\;\;\;{\bf k}\otimes 1 \otimes {\bf a}^-$}}

\put(285,0){\put(-22,24){$+$}
\put(10,40){\color{red}\line(1,0){40}} \put(50,40){\color{red}\vector(0,1){10}}
\put(0,20){\color{red}\line(1,0){30}} \put(30,20){\color{red}\vector(0,1){30}}
\put(0,0){\color{red}\line(1,0){10}}
\put(10,0){\color{red}\line(0,1){40}}
\put(0,40){\line(1,0){10}}
\put(10,40){\vector(0,1){10}}
\put(-22,-20){$+\;\; {\bf a}^-\otimes 1 \otimes 1$}}

\put(380,0){\put(-25,24){$+$}
\put(0,40){\color{red}\line(1,0){50}} \put(50,40){\color{red}\vector(0,1){10}}
\put(0,20){\color{red}\line(1,0){30}} \put(30,20){\color{red}\vector(0,1){30}}
\put(0,0){\color{red}\line(1,0){10}}
\put(10,0){\color{red}\vector(0,1){50}}
\put(-23,-20){$+\;\;\;1\otimes 1 \otimes 1$}}

\end{picture}
\end{equation*}
Here the edges are colored black or red depending on whether they assume 
$0$ or $1$, and the three components in the tensor product are 
arranged so as to correspond to the 
vertices at $(\text{top left}) \otimes (\text{bottom left})  \otimes 
(\text{top right}) $.
See (\ref{5v}).

Again general $n$ case is similar and clear.
By imposing the boundary condition according to (\ref{Lin}), 
the operator $X_i\,(0 \le i \le n)$ is given graphically as follows: 
\begin{equation}\label{Xn}
\begin{picture}(150,77)(-50,-10)
\put(-49,29){$X_i = \sum$}
\put(20,52){$. . .$}
\put(-5,27){$.$}\put(-5,24){$.$}\put(-5,21){$.$}
\put(-8,48){\line(1,0){56}}
\put(-8,40){\line(1,0){48}}
\put(-8,32){\line(1,0){40}}
\put(-8,16){\line(1,0){24}}
\put(-8,8){\line(1,0){16}}
\put(-8,0){\line(1,0){8}}

\put(11,9.5){\put(29,25){$.$}\put(27,23){$.$}\put(25,21){$.$}}
\put(-9,-9.5){\put(29,25){$.$}\put(27,23){$.$}\put(25,21){$.$}}

\put(48,48){\vector(0,1){8}}
\put(40,40){\vector(0,1){16}}
\put(32,32){\vector(0,1){24}}
\put(16,16){\vector(0,1){40}}
\put(8,8){\vector(0,1){48}}
\put(0,0){\vector(0,1){56}}

\put(51,46){$\scriptstyle{0}$}
\put(43,38){$\scriptstyle{0}$}
\put(30,25){$\scriptstyle{0}$}

\put(24,19){$\scriptstyle{1}$}
\put(11,6){$\scriptstyle{1}$}
\put(3,-2){$\scriptstyle{1}$}

\put(29,52){\rotatebox{-135}{$\overbrace{\phantom{KKKK}}$}}
\put(54,26){$\scriptstyle{n-i}$}

\put(2,23){\rotatebox{-135}{$\overbrace{\phantom{KKKk}}$}}
\put(26,-2){$\scriptstyle{i}$}

\end{picture}
\end{equation}
There are $n$ incoming arrows from the left and $n$-outgoing ones from the top.
It represents an element of ${\mathscr A}_0^{\otimes n(n-1)/2}$.
The formula (\ref{ptr}) remains the same if the trace is understood as the one 
over $F^{\otimes n(n-1)/2}$.
In general it is normalized so that 
$\mathbb{P}(\sigma_1,\ldots, \sigma_L) = 1$ if 
$\sigma_1\ge \cdots \ge \sigma_L$ (which means 
$\sigma_1=n, \sigma_L=0$ as we are in a basic sector).
As another example, the configurations contributing to 
$X_{1}$ with $n=4$ are as follows:

 \begin{center} 
 \includegraphics[height=1.5cm,width=13.5cm]{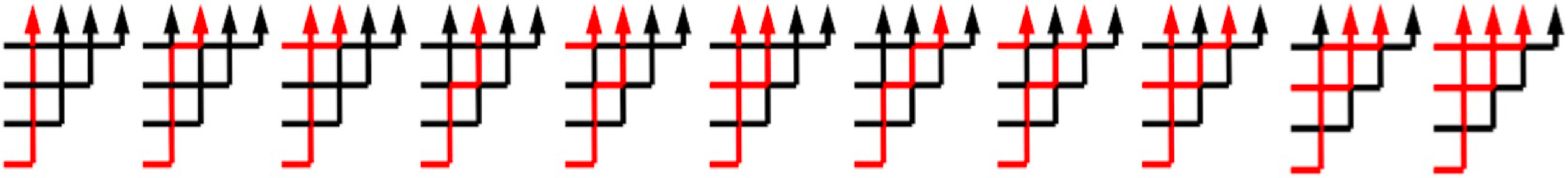}
 \end{center}
Accordingly we have
$X_1={\bf{k}}{\cdot}1{\cdot}1{\cdot}{\bf{k}}{\cdot}1{\cdot}{\bf{k}}+{\bf a}^{-}{\cdot}{\bf{a}}^{+}{\cdot}1{\cdot}{\bf k}{\cdot}1{\cdot}{\bf k}+1{\cdot}{\bf a}^{+}{\cdot}1{\cdot}{\bf k}{\cdot}1{\cdot}{\bf k}+1{\cdot}{\bf k}{\cdot}1{\cdot}{\bf a}^{-}{\cdot}{\bf a}^{+}{\cdot}{\bf k}+{\bf a}^{+}{\cdot}{\bf k}{\cdot}1{\cdot}{\bf a}^{-}{\cdot}{\bf a}^{+}{\cdot}{\bf k}+{\bf k}{\cdot}{\bf k}{\cdot}1{\cdot}1{\cdot}{\bf a}^{+}{\cdot}{\bf k}+1{\cdot}{\bf a}^{-}{\cdot}{\bf a}^{+}{\cdot}{\bf a}^{-}{\cdot}{\bf a}^{+}{\cdot}{\bf k}+{\bf a}^{+}{\cdot}{\bf a}^{-}{\cdot}{\bf a}^{+}{\cdot}{\bf a}^{-}{\cdot}{\bf a}^{+}{\cdot}{\bf k}+{\bf k}{\cdot}{\bf a}^{-}{\cdot}{\bf a}^{+}{\cdot}1{\cdot}{\bf a}^{+}{\cdot}{\bf k}+{\bf a}^{-}{\cdot}1{\cdot}{\bf a}^{+}{\cdot}1{\cdot}{\bf a}^{+}{\cdot}{\bf k}+1{\cdot}1{\cdot}{\bf a}^{+}{\cdot}1{\cdot}{\bf a}^{+}{\cdot}{\bf k}$,
where ${\cdot}$ denotes ${\otimes}$ and 
${\mathscr A}_0^{\otimes 6}$ is ordered as
$(\text{top line})\otimes (\text{middle line}) \otimes (\text{bottom line})$
where within each line they are ordered from the left to the right.
 
To summarize, our $X_i$ is a corner transfer matrix of the 
$q=0$-oscillator valued five-vertex model.
The steady state probability $\mathbb{P}({\boldsymbol \sigma})$ 
(\ref{ptr}) is a 
{\em partition function} of the 3D lattice model 
associated with the 3D $L$ operator at $q=0$ (\ref{5v}).
It is a 3D system having the prism shape region 
$(\ref{Xn}) \times \Z_L$
with the boundary condition on one of the surfaces 
specified by ${\boldsymbol \sigma}$.
The $X_i$ plays the role of a layer-to-layer transfer matrix.
In general it does not obey the recursion relation 
with respect to $n$ of the form
$X^{(n)}_i = \sum_{0 \le j \le n-1}a^{(n)}_{i,j} \otimes X^{(n-1)}_j$
for some $a^{(n)}_{i,j} \in \mathscr{A}_0^{\otimes n-1}$.
As a result it is different from those in \cite{EFM,AAMP} except 
one particular case of $n=3$ in the latter and 
the simplest nontrivial case $n=2$:
\begin{align*}
X_0 = 1+ {\bf a}^+,\;\;
X_1 = {\bf k},\;\;
X_2 = 1+{\bf a}^-.
\end{align*}

\section{Discussion}\label{sec:dis}

\subsection{Summary}
In this paper we have revealed that the Ferrari-Martin algorithm for 
constructing the steady state of the $n$-TASEP \cite{FM} is most naturally 
formulated in terms of a combinatorial $R$ in crystal base theory.
Combined with the factorized form of the combinatorial $R$
originating in the tetrahedron equation \cite{KOS}, it has lead to
a new matrix product formula (\ref{ptr}) for the steady state probability.
Our operator $X_i$ (\ref{Xn}) is a corner transfer matrix of the 
$q$-oscillator valued five-vertex model at $q=0$, which is 
a configuration sum in a cross-channel of the original problem.
Whether such a result admits generalizations to 
similar systems like ASEP \cite{PEM}, 
open boundary conditions
and the large list of factorized $R$ matrices 
for other quantum groups \cite{KOS}
is a natural question to be investigated.

\subsection{Generalized dynamics}\label{ss:gd} 
Let us comment on a possible generalization of the time evolutions
$T_i$ and  $\tau_i$ in Proposition \ref{pr:time} from the 
viewpoint of crystal base theory.
Let $B^{l,s}\,(1 \le l <L, s \ge 1)$ be the crystal of the 
irreducible $U_q(\widehat{sl}_L)$ module 
(so-called Kirillov-Reshetikhin module) corresponding to the 
$l \times s$ rectangular shape Young diagram \cite{KMN2,Sh}.
Elements of $B^{l,s}$ are labeled with the semi-standard 
tableaux on the Young diagram with entries $\{1,\ldots, L\}$. 
The family $B^{l,s}$ includes $B^l$ (\ref{VBdef}) as $B^{l,1}=B^l$,  
and again there is a bijection 
$B^{l,s} \otimes B^{l',s'} \rightarrow B^{l',s'} \otimes B^{l,s}$ 
called the combinatorial $R$.
For an arbitrary pair of $(l,s)$ and $(l',s')$,
an algorithm for finding the image of the $R$ 
is known \cite{Sh} which reduces to the NY-rule for $s=s'=1$.

Associated with any crystal $B^{r,s}$ 
we introduce the maps
${\mathcal T}={\mathcal T}^{r,s}: 
{\mathcal B}({\bf m}) \otimes B^{r,s}
\rightarrow B^{r,s} \otimes {\mathcal B}({\bf m})$
and ${\mathcal T}_u: {\mathcal B}({\bf m})\rightarrow {\mathcal B}({\bf m})$
for each $u \in B^{r,s}$ by a scheme similar to (\ref{T}) and (\ref{Tk}).
Namely we set 
${\mathcal T}({\bf b}_1 \otimes \cdots \otimes {\bf b}_n \otimes u)
= u' \otimes {\bf b}'_1 \otimes \cdots \otimes {\bf b}'_n$
and ${\mathcal T}_u({\bf b}_1 \otimes \cdots \otimes {\bf b}_n)
={\bf b}'_1 \otimes \cdots \otimes {\bf b}'_n$,
where $u'\in B^{r,s}$ and ${\bf b}'_i \in B^{l_i}$ 
are obtained by sending $u \in B^{r,s}$ to the left 
through ${\bf b}_1 \otimes \cdots \otimes {\bf b}_n$ 
by a successive application of the combinatorial $R$: 
$B^{l_i} \otimes B^{r,s} \rightarrow B^{r,s} \otimes B^{l_i}$.
Maps like ${\mathcal T}_u$ have been studied extensively
as a time evolution with {\em carrier} $u$ 
in the context of integrable cellular automata known as 
{\em box-ball systems} and their generalizations \cite{IKT}. 
Based on computer experiments we propose
\begin{conjecture}\label{cj:T}
For ${\boldsymbol \sigma} \in S({\bf m})$, 
let $\pi^{-1}({\boldsymbol \sigma}) \subset {\mathcal B}({\bf m})$ 
denote the preimage of the map $\pi$ (\ref{tpdef}).
Then the set 
$\pi \circ {\mathcal T}_u\bigl(\pi^{-1}({\boldsymbol \sigma})\bigr)$
consists of a single element for any $u \in B^{r,s}$.
\end{conjecture}
Admitting the conjecture we are able to introduce 
a new time evolution $\tau_u$ on $S({\bf m})$ by
$\tau_u := \pi \circ {\mathcal T}_u \circ \pi^{-1}$.
The definition is equivalent to 
replacing $T_i$ and $\tau_i$ in 
the commutative diagram of Proposition \ref{pr:time}
by ${\mathcal T}_u$ and $\tau_u$, respectively.
The detail of the dynamics will be illustrated elsewhere.
The map ${\mathcal T}=
{\mathcal T}^{r,s}$ is invertible as it is a composition of combinatorial $R$'s.

Now we introduce new stochastic processes
on ${\mathcal B}({\bf m})$ and $S({\bf m})$ analogous to 
the multiline process and the $n$-TASEP, respectively.
They are defined by letting  
each state undergo the respective time evolutions ${\mathcal T}_u$ and $\tau_u$
with an equal probability with respect to $u \in B^{r,s}$.
Then from the invertibility of ${\mathcal T}$, exactly the same 
argument as \cite[p821]{FM} proves that
the steady states are given by 
the uniform probability distribution for the former and by 
(\ref{pbar}) for the latter.
In this way we find a curious feature of 
the probability distribution (\ref{pbar}) 
that it yields a {\em simultaneous} steady state for 
the family of stochastic processes on $S({\bf m})$ associated with the crystals 
$B^{r,s}$ as well as for the $n$-TASEP.

\subsection{Hat relations}\label{ss:hat}
Our derivation of (\ref{ptr})  and  (\ref{Xn}) in this paper  
is based on Proposition \ref{pr:fm} which is attributed to \cite{FM}.
However it is also possible to establish them directly following the idea in \cite{DEHP}.  
To explain it 
let us write the action of the local Hamiltonian (\ref{H}) as 
$h |\alpha, \beta \kt = \sum_{0 \le \gamma,\delta\le n}
h^{\gamma,\delta}_{\alpha,\beta} |\gamma,\delta\kt$.
Then for the quantity 
$\mathrm{Tr}\bigl(X_{\sigma_1} \cdots X_{\sigma_L}\bigr)$
in (\ref{ptr}) to give the steady state probability, 
it is sufficient to construct the operators 
$\hat{X}_0,\ldots, \hat{X}_n \in \mathrm{End}(F^{\otimes n(n-1)/2})$ 
satisfying the so called hat relation \cite{DEHP}:
\begin{align}\label{hatr}
X_\alpha\hat{X}_\beta - \hat{X}_\alpha X_\beta 
= (hXX)_{\alpha, \beta} 
:= \sum_{0 \le \gamma,\delta \le n}
h^{\alpha,\beta}_{\gamma,\delta}\,X_\gamma X_\delta\quad
(0 \le \alpha, \beta \le n).
\end{align}
\begin{theorem}\label{th:hat}
Up to a freedom $\hat{X}_i \rightarrow \hat{X}_i + 
\mathrm{const}\,X_i$ at least, such an $\hat{X}_i$ is realized as 
\begin{equation}\label{Xhn}
\begin{picture}(150,77)(-70,-10)
\put(-115,29){$\hat{X}_i = \sum(\alpha_1+\cdots+ \alpha_n)$}
\put(20,52){$. . .$}
\put(-5,27){$.$}\put(-5,24){$.$}\put(-5,21){$.$}
\put(-8,48){\line(1,0){56}}
\put(-8,40){\line(1,0){48}}
\put(-8,32){\line(1,0){40}}
\put(-8,16){\line(1,0){24}}
\put(-8,8){\line(1,0){16}}
\put(-8,0){\line(1,0){8}}

\put(11,9.5){\put(29,25){$.$}\put(27,23){$.$}\put(25,21){$.$}}
\put(-9,-9.5){\put(29,25){$.$}\put(27,23){$.$}\put(25,21){$.$}}

\put(48,48){\vector(0,1){8}}\put(44,60){$\scriptstyle{\alpha_n}$}
\put(40,40){\vector(0,1){16}}
\put(32,32){\vector(0,1){24}}
\put(16,16){\vector(0,1){40}}
\put(8,8){\vector(0,1){48}}\put(5,60){$\scriptstyle{\alpha_2}$}
\put(0,0){\vector(0,1){56}}\put(-7,60){$\scriptstyle{\alpha_1}$}

\put(51,46){$\scriptstyle{0}$}
\put(43,38){$\scriptstyle{0}$}
\put(30,25){$\scriptstyle{0}$}

\put(24,19){$\scriptstyle{1}$}
\put(11,6){$\scriptstyle{1}$}
\put(3,-2){$\scriptstyle{1}$}

\put(29,52){\rotatebox{-135}{$\overbrace{\phantom{KKKK}}$}}
\put(54,26){$\scriptstyle{n-i}$}

\put(2,23){\rotatebox{-135}{$\overbrace{\phantom{KKKk}}$}}
\put(26,-2){$\scriptstyle{i}$}

\end{picture}
\end{equation}
where the sum ranges over the same set of configurations of the 
five-vertex model as in $X_i$ (\ref{Xn}).
The only difference is the extra coefficient 
$\alpha_1+\cdots+ \alpha_n$ depending on the $n$ outgoing arrows 
$\alpha_i = 0, 1$.
\end{theorem}

For instance one has 
$\hat{X}_0 = 
{\bf a}^+ \otimes 1 \otimes 1 +
{\bf k} \otimes {\bf a}^+  \otimes 1 +
{\bf a}^- \otimes {\bf a}^+ \otimes {\bf a}^+ +
2(1 \otimes {\bf a}^+ \otimes {\bf a}^+)$
for $n=3$ from the graphical representation of 
$X_0$ in Section \ref{ss:mpr}.
The hat relation (\ref{hatr}) turns out to be 
a consequence of its far-reaching generalization
into a 3D lattice model obeying the tetrahedron equation.
An exposition of the detail is beyond the scope of this paper and 
will be presented in \cite{KMO}  
together with a proof of Theorem \ref{th:hat}.

\section*{Acknowledgments}
The authors thank 
Yasuhiko Yamada for useful discussion and Chikashi Arita for a communication.
A.~K. thanks 
Tetsuo Deguchi, Akinori Nishino, Tomohiro Sasamoto and Hal Tasaki
for kind interest and valuable comments at  
Mathematical and Statistical Physics Seminar at
Ochanomizu University, 6 June 2015, where 
a part of the work was presented.
This work is supported by 
Grants-in-Aid for Scientific Research No.~15K04892,
No.~15K13429 and No.~23340007 from JSPS.

\end{document}